\documentclass[]{aa}

\usepackage{graphicx}
\usepackage{natbib}
\usepackage{scalerel}

\usepackage[table]{xcolor}

\usepackage{txfonts}
\usepackage[pdfencoding=auto,psdextra]{hyperref}
\hypersetup{
    colorlinks=true,
    linkcolor=blue,
    filecolor=magenta,      
    urlcolor=blue,
    citecolor=blue
}
\makeatletter
\renewcommand*\aa@pageof{, page \thepage{} of \pageref*{LastPage}}
\makeatother

\usepackage[utf8]{inputenc}

\usepackage[switch, modulo]{lineno}

\usepackage{euclid}
\usepackage{physics}

\defcitealias{Singh2015}{S+15}
\defcitealias{Singh2016}{S+16}

\begin{document}

\newcommand{\eps}{\epsilon} 
\newcommand{\epsI}{\epsilon_\mathrm{I}} 
\newcommand{\shear}{\gamma} 

\newcommand{\epsChi}{\epsilon}
\newcommand{\epsIChi}{\epsilon_\mathrm{I}}
\newcommand{\shearChi}{\gamma}

\newcommand{\tang}[1]{#1^\mathrm{t}} 
\newcommand{\fourier}[1]{\Tilde{#1}} 

\newcommand{\convWL}{\kappa} 
\newcommand{\convIA}{\kappa_\mathrm{I}} 
\newcommand{\convWLChi}{\kappa} 
\newcommand{\convIAChi}{\kappa_\mathrm{I}} 
\newcommand{\deltaMatter}{\delta} 
\newcommand{\deltaIA}{\delta_\mathrm{I}} 
\newcommand{\deltaGal}{\delta_\mathrm{g}} 

\newcommand{\numDenThreeD}{n} 
\newcommand{\aveNumDenThreeD}{\Bar{n}}
\newcommand{\numDenTwoD}{N}
\newcommand{\aveNumDenTwoD}{\Bar{N}}

\newcommand{\GGGL}{G} 
\newcommand{\GGGLShear}{G_{\mathrm{gg}\gamma}}
\newcommand{\GGGLIA}{G_{\mathrm{ggI}}}

\newcommand{\Pimax}{\Pi_\mathrm{m}}
\newcommand{\Nap}{N_\mathrm{ap}}
\newcommand{\Map}{M_\mathrm{ap}} 
\newcommand{\MapIA}{M_\mathrm{ap}^\mathrm{I}} 

\newcommand{\NNMIA}{\expval{\Nap\Nap\MapIA}}

\newcommand{\varthetavec}{\vec{\vartheta}}
\newcommand{\ellvec}{\vec{\ell}}
\newcommand{\I}{\mathrm{i}}
\newcommand{\E}{\mathrm{e}}
\newcommand{\thetavec}{\vec{\theta}}
\newcommand{\rvec}{\vec{r}}
\newcommand{\kvec}{\vec{k}}

\newcommand{\dirac}{\delta_\mathrm{D}}
%
%

\title{Third-order intrinsic alignment of SDSS BOSS LOWZ galaxies}

   
\newcommand{\orcid}[1]{} 
\author{\normalsize Laila~Linke$^{1,2}$\thanks{Corresponding author,
    \email{laila.linke@uibk.ac.at}}, Susan~Pyne$^{3}$, Benjamin~Joachimi$^{3}$, Christos~Georgiou$^{4}$, Kai~Hoffmann$^{5}$, Rachel~Mandelbaum$^{6}$, Sukhdeep~Singh$^{6}$}

%
%
%
%

\institute{$^{1}$ Universit\"at Innsbruck, Institut f\"ur Astro- und Teilchenphysik, Technikerstr. 25/8, 6020 Innsbruck, Austria\\
$^{2}$ Universit\"at Bonn, Argelander-Institut f\"ur Astronomie, Auf dem H\"ugel 71, 53121 Bonn, Germany\\ 
$^{3}$ Department of Physics and Astronomy, University College London, Gower Street, London WC1E 6BT, UK\\
$^{4}$ Institute for Theoretical Physics, Utrecht University, Princetonplein 5, 3584 CC, Utrecht, The Netherlands.\\
$^{5}$ Institute of Space Sciences (ICE, CSIC), Campus UAB, Carrer de Can Magrans, s/n, 08193 Barcelona, Spain \\
$^{6}$ McWilliams Center for Cosmology \& Astrophysics, Department of Physics, Carnegie Mellon University, Pittsburgh, PA 15213, USA}
%
%
\abstract{Cosmic shear is a powerful probe of cosmology, but it is affected by the intrinsic alignment (IA) of galaxy shapes with the large-scale structure. Upcoming surveys such as \textit{Euclid} and Vera C. Rubin Observatory's Legacy Survey of Space and Time (LSST) require an accurate understanding of IA, particularly for higher-order cosmic shear statistics that are vital for extracting the most cosmological information. In this paper, we report the first detection of third-order IA correlations using the LOWZ galaxy sample from the Sloan Digital Sky Survey (SDSS) Baryon Oscillation Spectroscopic Survey (BOSS). We compare our measurements with predictions from the MICE cosmological simulation and an analytical model inspired by the Non-linear Linear Alignment (NLA) model and informed by second-order correlations. We also explore the dependence of the third-order correlation on the galaxies' luminosity. We find that the amplitude $A_\mathrm{IA}$ of the IA signal is non-zero at the $4.7\sigma$ ($7.6\sigma$) level for scales between $6\,h^{-1}\,\mathrm{Mpc}$ ($1\,h^{-1}\,\mathrm{Mpc}$) and $20\,h^{-1}\,\mathrm{Mpc}$. For scales above $6\,h^{-1}\,\mathrm{Mpc}$ the inferred $A_\mathrm{IA}$ agrees both with the prediction from the simulation and estimates from second-order statistics within 1$\sigma$ but deviations arise at smaller scales. Our results demonstrate the feasibility of measuring third-order IA correlations and using them for constraining IA models. The agreement between second- and third-order IA constraints also opens the opportunity for a consistent joint analysis and IA self-calibration, promising tighter parameter constraints for upcoming cosmological surveys.}
%
%
%
%
   \titlerunning{ Third-order IA in LOWZ}
   \authorrunning{ Linke et al.}
   
   \maketitle
%
%
%
%
   
\section{\label{sc:Intro}Introduction}
Cosmic shear, the correlations between galaxy shapes due to the weak gravitational lensing by the cosmic large-scale structure, is a popular and constraining probe of cosmology. Cosmic shear surveys such as the Dark Energy Survey (DES, \citealp{Abbott2016, Becker2016}), the Kilo-Degree Survey (KiDS, \citealp{Kuijken2015}), or the Hyper-Suprime-Cam survey (HSC, \citealp{Aihara2018}) constrain the parameter $S_8=\sigma_8\,\sqrt{\Omega_\mathrm{m}/0.3}$, which is a combination of the matter density parameter $\Omega_\mathrm{m}$ and the clustering parameter $\sigma_8$, at $2$--$4$ per cent \citep{Asgari2021, Amon2022, Secco2022, Dalal2023, Li2023}. Soon, even more extensive and deeper surveys such as \textit{Euclid} \citep{Laureijs2011, EuclidSkyOverview}, Vera C. Rubin Observatory's Legacy Survey of Space and Time (LSST, \citealp{Ivezic2019}) or the Nancy Grace Roman space telescope \citep{Akeson2019} will use cosmic shear in combination with the correlations of galaxy positions to not only better constrain $\Omega_\mathrm{m}$ and $\sigma_8$, but also determine the time evolution of dark energy \citep{LSST2018, Blanchard-EP7}. Arguably, cosmic shear is now entering its most exciting phase as a probe of our Universe.

While the increasing size of the data sets provides more and more precision, the accuracy of theoretical predictions also needs to be updated. For this, several astrophysical effects that impact cosmic shear measurements must be understood. One such effect is the alignment between the intrinsic shapes of galaxies, the so-called intrinsic alignment (IA, \citealp{Lamman2024, Troxel2015, Joachimi2015}). IA occurs due to the formation of galaxies within the dark matter-dominated gravitational field \citep{Catelan2001}. Elliptical galaxies tend to be aligned with the gravitational tidal field, leading to a radial alignment with respect to matter overdensities \citep{Kiessling2015}. These alignments add to the observed shape correlations and thus contaminate the cosmic shear signal.

If IA is not taken into account in cosmic shear analyses, inferred parameters can be severely biased \citep{Hirata2007, Kirk2015}. Therefore, cosmic shear analyses commonly incorporate a description for IA in their model with some free parameters that are constrained in the inference simultaneously with the cosmological parameters. 

Several IA models have been proposed for this purpose. A relatively simple model is the non-linear linear alignment (NLA) model by \citet{Bridle2007}. This empirical model assumes that the power spectrum of galaxy shapes is linearly related to the non-linear matter power spectrum via a free parameter $A_\mathrm{IA}$ describing the amplitude of the correlation. While it is only a phenomenological description, it fits well to IA measurements for red central galaxies at large scales \citep{Singh2015}. There are also other more physically motivated models, such as the Tidal Alignment and Tidal Torque (TATT) model \citep{Blazek2019}, which allows galaxies to not align solely with the gravitational tidal field but also the tidal torque and includes a density weighting term. Alternatively, there are halo model descriptions \citep{Schneider2010, Fortuna2021}, which explicitly distinguish between central and satellite galaxies inside dark matter halos. Other models are based on perturbative approaches \citep{Vlah2020, Maion2024, Chen2024}. The accuracy of the models depends on the considered physical scales and galaxy samples. It is currently unclear which IA models offer the best trade-off between model complexity and accuracy for any given analysis. For example, there is no clear observational evidence yet on whether the more complex TATT or the simpler NLA model are more appropriate for cosmological analyses \citep{DES_KiDS2023}.

However, the choice of model can impact cosmological parameter constraints. \citet{DES_KiDS2023} find in a combined analysis of DES and KiDS that using the TATT instead of the NLA model can lower the inferred value for $S_8$ by almost 3 per cent, corresponding to a shift by $0.9 \sigma$. Consequently, it is paramount to test IA models independently of cosmic shear measurements to find an appropriate model. Furthermore, independent constraints of IA model parameters can be used as priors for cosmic shear analyses, significantly tightening constraints on other parameters \citep{Johnston2019}.


IA studies so far have been mainly concerned with second-order correlations of matter overdensity and intrinsic galaxy shapes. Mostly the shape-shape and density-shape correlations are constrained \citep[e.g.][]{Joachimi2011, Singh2015, Singh2016, Johnston2019, Fortuna2021, Johnston2021, Samuroff2023}. For several reasons, though, it has become increasingly interesting to study higher-order statistics (HOS), for example third-order correlations. For example, while current IA models are well-designed to describe second-order IA correlations, it is an important consistency check to see whether they also match higher-order correlations. An example of such a test is carried out by \citet{Pyne2022}, who measure the IA power-  and bispectrum in simulations and find that the same model parameters could describe both. Furthermore, HOS depend differently on both cosmological and IA parameters. Combining second-order and HOS can reduce parameter degeneracies and lead to better constraints \citep[e.g.][]{Burger2024, Ajani-EP29}. Adding higher-order information can also help in self-calibrating systematic effects such as IA \citep{Pyne2021}. Analyses similar to these are vital to optimally use the data sets of \textit{Euclid} and LSST, but they can only be performed if we can ensure we understand how systematic effects such as IA behave for third-order statistics. As shown by \citet{Semboloni2008} using cosmological simulations, third-order correlations of intrinsic galaxy shapes and the matter field can significantly impact analysis of third-order cosmic shear and contribute as much as 15\% of the signal for surveys with a median redshift of 0.7. Consequently, we must include IA in the modelling of the third-order cosmic shear signal and demonstrate that our models can accurately describe higher-order correlations of the matter overdensity and the intrinsic shapes of galaxies.

In this paper, we take the first step towards this by reporting the first detection of third-order intrinsic alignment correlations in the low-redshift (LOWZ) galaxy sample from the Sloan Digital Sky Survey (SDSS) Baryon Oscillation Spectroscopic Survey (BOSS) survey. This sample contains spectroscopically observed luminous red galaxies (LRGs) at redshifts below 0.4. For this sample, \citet{Singh2015} measured second-order correlations between the matter density and the shapes and between shapes of different galaxies. The measurement was conducted by using the positions of galaxies as tracers for the matter distribution and comparing them pairwise to the galaxies' shapes. We followed in their footsteps by comparing the positions of galaxy pairs with the shape of a third galaxy, thus measuring the density-density-shape correlation. This correlation was then compared to predictions by the Marenostrum Institut de Ciències de l'Espai (MICE) cosmological simulation and an analytical NLA-based model.  

This paper is structured as follows. In Sect.~\ref{sc:Theo}, we define fundamental notations, describe the third-order intrinsic alignment correlation function and relate it to the matter-matter-shape bispectrum. In Sect.~\ref{sc:Measurement}, we describe our estimator for the correlation function and our covariance estimate. Section~\ref{sc:Data} describes the observed and simulated data sets. The measurements and comparisons to the analytical model are presented in Sect.~\ref{sc:Results}. We conclude with a discussion in Sect.~\ref{sc:Discussion}.

\section{\label{sc:Theo} Theoretical background and modelling  }
\subsection{Basic quantities and definitions}

In the weak gravitational lensing limit (see e.g. \citealp{Bartelmann2001} for a review), the observed ellipticity $\eps(\varthetavec)$ of a source galaxy at angular position $\varthetavec$ is determined by its intrinsic ellipticity $\epsI(\varthetavec)$ and the weak lensing shear $\shear(\varthetavec)$, both of which are complex quantities. In weak lensing, $|\gamma|\ll 1$, and,
\begin{equation}
\label{eq: ellipticity}
    \eps(\varthetavec)=\epsI(\varthetavec)+\shear(\varthetavec)\;.
\end{equation}
Under the flat sky approximation, the shear is related to the lensing convergence $\convWL$ via the Kaiser-Squires relation \citep{Kaiser1993}
\begin{equation}
    \fourier{\shear}(\ellvec) = \E^{2\I\,\phi_\ell} \fourier{\convWL}(\ellvec)\;,
\end{equation}
where the tilde denotes Fourier transform and $\phi_\ell$ is the polar angle of the angular frequency $\ellvec$. Analogously, we can define an `intrinsic alignment (IA) convergence' $\convIA$, which describes the convergence that would cause a shear equivalent to the intrinsic source ellipticity via
\begin{equation}
    \fourier{\epsI}(\ellvec) = \E^{2\I\,\phi_\ell} \fourier{\convIA}(\ellvec)\;.
\end{equation}
For source galaxies distributed with a probability distribution $p(\chi)$ with comoving distance $\chi$, $\convWL(\varthetavec)$ is a weighted integral over the $\convWLChi(\varthetavec, \chi)$ at each $\chi$,
\begin{align}
\label{eq: definition convergence}
    \convWL(\varthetavec) &= \int \dd{\chi} p(\chi)\, \convWLChi(\varthetavec, \chi) \\
    &\notag = \int \dd{\chi} p(\chi)\, \int \dd{\chi'} W(\chi, \chi') \, \deltaMatter(\chi'\varthetavec, \chi')\;,
\end{align}
where we assumed a flat Universe, $\deltaMatter(\chi\varthetavec, \chi)$ is the matter density contrast at angular position $\varthetavec$ and distance $\chi$, and $W$ is the lensing efficiency given by
\begin{equation}
    W(\chi, \chi') = \frac{3\,H_0^2\,\Omega_\mathrm{m}}{2c^2}\, \frac{\chi' (\chi-\chi')}{\chi\, a(\chi')}\;,
\end{equation}
where $H_0$ is the Hubble constant, $\Omega_\mathrm{m}$ is the matter density parameter, and $a(\chi)$ is the scale factor at co-moving distance $\chi$. The $\convIA$ can be related to a density contrast $\deltaIA$, which would cause a shear equivalent to the intrinsic galaxy ellipticity,
\begin{equation}
    \convIA(\varthetavec) = \int \dd{\chi} p(\chi)\, \deltaIA(\chi\varthetavec, \chi)\;.
\end{equation}

We are interested in correlations between the intrinsic ellipticity field $\deltaIA$ and the matter density contrast $\deltaMatter$. To study these correlations, we use galaxy positions as tracers of the matter field. Their distribution is characterised by their three-dimensional number density $\numDenThreeD(\chi \varthetavec, \chi)$ at angular position $\varthetavec$ and distance $\chi$. This density is related to the galaxy number density contrast $\deltaGal$ by
\begin{equation}
    \numDenThreeD(\chi\varthetavec, \chi) = \aveNumDenThreeD(\chi) \left[\deltaGal(\chi\varthetavec, \chi)+1\right]\;,
\end{equation}
where $\aveNumDenThreeD(\chi)$ is the average number density at comoving distance $\chi$. From this, we also define the projected, two-dimensional galaxy number density $\numDenTwoD(\varthetavec)$ as
\begin{equation}
\label{eq: definition projected number density}
    \numDenTwoD(\varthetavec) = \int \dd{\chi} p(\chi)\, \numDenThreeD(\chi\varthetavec, \chi)\;,
\end{equation}
whose average we denote as $\aveNumDenTwoD$.

\subsection{Intrinsic alignment correlation functions}

Since the intrinsic ellipticity of an individual galaxy is unknown, its observed ellipticity alone does not contain usable cosmological information. Instead, one needs to consider the correlation functions of the ellipticities. The most commonly used ones are the second-order correlations, either of the ellipticities $\epsilon$ with themselves (i.e. `cosmic shear') or between ellipticities and the galaxy number density $\numDenTwoD$ (i.e. `galaxy-galaxy-lensing'). Here, though, we are interested in third-order correlation functions involving galaxy shapes. For this, we have, in principle, three choices: correlating the ellipticities at three different positions (`third-order cosmic shear', \citealp{Schneider2003})), the ellipticities at two positions with the galaxy number density at a third position, or the ellipticity at one position with the number density at two different positions (`galaxy-galaxy-galaxy lensing', G3L, \citealp{Schneider2005}). The last of these, so-called shape-lens-lens G3L, has the highest signal-to-noise ratio \citep{Simon2012}, so we concentrate on this.

Additionally, one can also construct the correlation between the galaxy number density at three different positions (`third-order galaxy clustering') \citep{Scoccimarro2001, Takada2003}. While this correlation does not include any information on the shapes or IA of the galaxies, it is a useful statistic of the cosmic large-scale structure, has been measured in several galaxy surveys \citep[e.g][]{Marin2013, Slepian2017} and can improve constraints from cosmological inference \citep{Yankelevich2019, Eggemeier2021, Ivanov2023}

In general, the correlation function for shape-lens-lens G3L is
\begin{equation}
    \GGGL'(\varthetavec_1, \varthetavec_2)  = \frac{1}{\aveNumDenTwoD^2}\expval{\numDenTwoD(\thetavec+\varthetavec_1)\,\numDenTwoD(\thetavec+\varthetavec_2)\, \eps(\thetavec)\, \E^{-\I(\phi_1+\phi_2)}}\;,
\end{equation}
where the $\varthetavec_i$ are the angular separations between the lens galaxies and the shape tracing galaxy, with polar angles $\phi_i$. This function can be decomposed into the sum of the contribution $\GGGLShear'$ from the shear and $\GGGLIA'$ from the intrinsic ellipticities as
\begin{equation}
    \GGGLShear'(\varthetavec_1, \varthetavec_2)  = \frac{1}{\aveNumDenTwoD^2}\expval{\numDenTwoD(\thetavec+\varthetavec_1)\,\numDenTwoD(\thetavec+\varthetavec_2)\, \shear(\thetavec)\, \E^{-\I(\phi_1+\phi_2)}}\;,
\end{equation}
and
\begin{equation}
    \GGGLIA'(\varthetavec_1, \varthetavec_2)  = \frac{1}{\aveNumDenTwoD^2}\expval{\numDenTwoD(\thetavec+\varthetavec_1)\,\numDenTwoD(\thetavec+\varthetavec_2)\, \epsI(\thetavec)\, \E^{-\I(\phi_1+\phi_2)}}\;.
\end{equation}

If the distances $\chi$ of the considered lens and source galaxies are known, one can also measure the correlation functions $\GGGL$, $\GGGLShear$, and $\GGGLIA$ in terms of the projected physical separations between the galaxies, for example,
\begin{align}
    \GGGLIA(\rvec_1, \rvec_2)  &= \frac{1}{\aveNumDenTwoD} \int \dd{\chi} p(\chi)\,\int \dd{\chi_1} p(\chi_1)\,\int \dd{\chi_2} p(\chi_2)\,\\
    &\notag\quad\times \expval{\numDenThreeD(\rvec+\rvec_1, \chi_1)\,\numDenThreeD(\rvec+\rvec_2, \chi_2)\, \epsI(\rvec, \chi)\, \E^{-\I(\phi_1+\phi_2)}}\;,
\end{align}
where $\epsI(\rvec, \chi)$ is the intrinsic shape of a galaxy at projected galaxy separation $\rvec$ on the sky (in physical units) and distance $\chi$. The functions $\GGGL$ and $\GGGLShear$ can be defined analogously by replacing the intrinsic shape with the total observed shape or the shear.

Usually, when measuring the G3L correlation function, one uses lens and source samples that are well separated along the line of sight \citep{Simon2012, Linke2020b}. This increases $\GGGLShear$ since it depends on the correlation between the matter structures in front of the sources and the lens number density, weighted by the lensing efficiency $W$. For a given lens distance, the efficiency is stronger if the sources are farther away from the lensing structures and, thus, also the lens galaxies. The intrinsic contribution $\GGGLIA$, though, is down-weighted if lenses and sources are far apart from each other. This is because the intrinsic ellipticity of the sources depends primarily on their local density distribution. Therefore, the correlation to the lens number density is strongest if sources and lenses are at the same co-moving distance.
Here, we are particularly interested in $\GGGLIA$, so we reverse the usual process of using well-separated lens and source samples. Instead, we explicitly measure the G3L signal only for lenses and sources with distances smaller than fixed physical distance $\Pimax$. Thus, using Eq.~\eqref{eq: definition projected number density}, the correlation function we measure is
\begin{align}
    &\notag\GGGL^{\Pimax}(\rvec_1, \rvec_2)\\
    &= \frac{1}{\aveNumDenTwoD^2_{\Pimax}} \int \dd{\chi} p(\chi) \int_{\chi-\Pimax}^{\chi+\Pimax} \dd{\chi_1} p(\chi_1)  \int_{\chi-\Pimax}^{\chi+\Pimax} \dd{\chi_2} p(\chi_2) \, \\
    &\notag \quad \times \expval{\numDenThreeD(\rvec+\rvec_1, \chi_1)\,\numDenThreeD(\rvec+\rvec_2, \chi_2)\, \epsChi(\rvec, \chi)\, \E^{-\I(\phi_1+\phi_2)}}\;,
\end{align}
where $\aveNumDenTwoD^2_{\Pimax}$ is the average number density of galaxy pairs within $\pm \Pimax$,
\begin{equation}
    \aveNumDenTwoD^2_{\Pimax} = \int \dd{\chi} p(\chi) \left[\int_{\chi-\Pimax}^{\chi+\Pimax} \dd{\chi_1} p(\chi_1) \, \aveNumDenTwoD(\chi_1)\right]^2\;.
\end{equation}
In principle, one can model $\GGGL^{\Pimax}$ for any $\Pimax$ by including both the IA and the cosmic shear contribution, as done for second-order statistics \citep{Samuroff2023}. Here, though, we choose a small $\Pimax$ such that the cosmic shear contribution is suppressed. This is possible because for a small enough $\Pimax$, $\chi_1$ and $\chi_2$ are close to $\chi$, so the lensing efficiency suppresses the correlation between $\numDenThreeD$ and the shear $\shear$. Therefore, we can replace $\epsChi$ by the intrinsic shape $\epsI$. Simultaneously, we choose $\Pimax$ large enough that all intrinsic alignment contributions to the correlation function are taken into account. Then, we can replace $\Pimax$ with infinity, so $\GGGL^{\Pimax}\rightarrow\GGGLIA$. We test whether this assumption holds in Sect.~\ref{sc:Results} using the simulated data (see Sect.~\ref{sc:Data}), which only includes IA and no cosmic shear contribution.


\subsection{Aperture statistics}
In the following, we are not directly analysing the correlation function but instead convert it to aperture statistics \citep{Schneider2002}. These have several practical advantages. For example, they can compress the data vector length from several hundred for the correlation function to a few tens. Moreover, they are more easily modelled from a bispectrum model than the correlation functions.


Aperture statistics are moments of aperture masses $\Map'$, 
\begin{align}
	\label{eq:DefinitionApertureMass}
	\Map'(\theta;\varthetavec)    &= \int \dd[2]{\vartheta'} U_\theta(|\varthetavec-\varthetavec'|)\,\convWL(\varthetavec')\,
 \end{align}
  and aperture number counts $\Nap'$,
  \begin{align}
	\Nap'(\theta;\varthetavec)    &= \frac{1}{\aveNumDenTwoD} \int \dd[2]{\vartheta'} U_\theta(|\varthetavec-\varthetavec'|)\,\numDenTwoD(\varthetavec')\;,
\end{align} 
where  $U_\theta$ is a compensated filter function $U_\theta$ of aperture scale radius $\theta$, that is $\int \dd{\vartheta}\, \vartheta\, U_\theta(\vartheta) = 0$. The $\Map'$ can be linked to the shear via
\begin{equation}
	\Map'(\theta;\varthetavec)    = \int \dd[2]{\vartheta'} Q_\theta(|\varthetavec-\varthetavec'|)\,\shear(\varthetavec') \;,
\end{equation}
where $Q$ is related to $U$ via
\begin{equation}
    Q_\theta(\vartheta) = \frac{2}{\vartheta^2}\,\int_0^\vartheta \dd{\vartheta'} \vartheta'\, U_\theta(\vartheta') - U_\theta(\vartheta)\;.
\end{equation}

We also define aperture measures $\Map$ and $\Nap$ as functions of projected physical separations $\rvec$ by
\begin{align}
    \Map(R; \rvec) &=\!\!\! \int\!\!\! \dd[2]{r'} \! U_R(|\rvec-\rvec'|)\!\! \!\int \!\!\!\dd{\chi}\! p(\chi)
    \!\!\!\int\! \!\! \dd{\chi'} W(\chi, \chi')\, \deltaMatter(\rvec', \chi')\\
    \Nap(R; \rvec) &= \int \dd[2]{r'} U_R(|\rvec-\rvec'|) \int \dd{\chi} p(\chi)\, \deltaGal(\rvec', \chi)\;.
\end{align}
Here, $R$ is the aperture scale radius, but in contrast to the $\theta$ in Eq.~\eqref{eq:DefinitionApertureMass}, it is in the same physical units, for example, Mpc, as $\rvec$. We also used the relations between three-dimensional densities and projected quantities from Eqs.~\eqref{eq: definition convergence} and~\eqref{eq: definition projected number density}.

In analogy to $\Map$, we define an intrinsic alignment aperture measure $\MapIA$ as
\begin{align}
    	\MapIA(R;\rvec)    &= \int \dd[2]{r'} U_R(|\rvec-\rvec'|) \int \dd{\chi} p(\chi)\, \deltaIA(\rvec', \chi)\;.
\end{align}
Throughout, we use the $U_\theta$  from \citet{Crittenden2002},
\begin{equation}
	\label{eq:exponentialFilterFunction}
	U_\theta(\vartheta)=\frac{1}{2\pi\theta} \, \left(1-\frac{\vartheta^2}{2\theta^2}\right)\, \exp(-\frac{\vartheta^2}{2\theta^2})\;.
\end{equation}
For this filter,  \citet{Schneider2005} showed that the correlation function $\GGGLShear'$ can be converted to the third-order aperture statistic through
\begin{align}
	\label{eq:NNMapFromGtilde}
&\notag	\expval{\Nap'\Nap'\Map'}(\theta_1, \theta_2, \theta_3)\\ 
&=\expval{\Nap'(\theta_1;\varthetavec) \Nap'(\theta_2;\varthetavec) \Map'(\theta_3;\varthetavec) }\\
 &\notag= \int \dd[2]{\vartheta_1} \, \int \dd[2]{\vartheta_2} \, \GGGLShear'(\varthetavec_1, \varthetavec_2)\, \mathcal{A}_{NNM}(\vartheta_1, \vartheta_2, \phi \mid \theta_1, \theta_2, \theta_3)\;.
\end{align}
where $\phi$ is the angle between $\varthetavec_1$ and $\varthetavec_2$ and the kernel function $\mathcal{A}_{NNM}$ is given in the appendix of \citet{Schneider2005}. 
Similarly, by transforming variables from $\varthetavec_i$ to $\rvec_i$,
\begin{align}
\label{eq: NNMIA from G}
   &\notag \expval{\Nap\Nap\MapIA}(R_1, R_2, R_3)\\
    &=\expval{\Nap(R_1; \rvec)\,\Nap(R_2; \rvec)\,\MapIA(R_3;\rvec)}\\
     &\notag= \int \dd[2]{r_1} \, \int \dd[2]{r_2} \, \GGGLIA(\rvec_1, \rvec_2)\, \mathcal{A}_{NNM}(r_1, r_2, \phi \mid R_1, R_2, R_3)\;,
\end{align}
Our main observable is the aperture statistics $\NNMIA$ for equal aperture radii $R_1=R_2=R_3=:R$. 
As shown in App.~\ref{app: bispectrum}, $\NNMIA$ can be related to the galaxy-galaxy-shape bispectrum $B_\mathrm{ggI}$ as
\begin{align}
\label{eq: NNMIA as bispec}
    &\notag\NNMIA(R, R, R)\\
    & =   \int \frac{\dd[2]{k_{\perp, 1}}}{(2\pi)^2}\int \frac{\dd[2]{k_{\perp, 2}}}{(2\pi)^2}\, \fourier{U}_R(k_{\perp, 1})\, \fourier{U}_R(k_{\perp, 2})\fourier{U}_R(|k_{\perp, 1}+k_{\perp, 2}|)\\
    &\notag \quad \times \int \dd{\chi} p^3(\chi) B_\mathrm{ggI}(k_{\perp, 1}, k_{\perp, 2}, k_{\perp, 3}; \chi, \chi, \chi)\;.
\end{align}

\subsection{Bispectrum model}
We model the bispectrum $B_\mathrm{ggI}$ motivated by the Non-linear alignment (NLA) IA model \citep{Hirata2004, Bridle2007}. According to this model, the power spectrum $P_{\delta\mathrm{I}}$ between matter densities and  intrinsic ellipticities is
\begin{equation}
    P_{\delta\mathrm{I}}(k) = -A_{\mathrm{IA}}\frac{C_1\Omega_\mathrm{m}\rho_\mathrm{cr}}{D(z)}\, P(k) = f_\mathrm{IA}\, P(k)\;,
\end{equation}
where $P$ is the non-linear matter power spectrum, $\Omega_\mathrm{m}$ is the matter density parameter, $\rho_\mathrm{cr}$ is the critical density, $D$ is the growth factor normalized to unity at the time the alignment is assumed to have occurred, $C_1$ is a normalization constant and $A_\mathrm{IA}$ is the intrinsic alignment amplitude. We extend this model to the bispectrum by assuming that the bispectrum $B_{\delta\delta\mathrm{I}}$ between matter densities and the intrinsic ellipticities is
\begin{equation}
    B_{\delta\delta\mathrm{I}}(k_1, k_2, k_3; \chi_1, \chi_2, \chi_3) = f_\mathrm{IA}\,  B(k_1, k_2, k_3; \chi_1, \chi_2, \chi_3)\;,
\end{equation}
where $B$ is the matter bispectrum. 
The bispectrum $B_\mathrm{ggI}$ between galaxies and the intrinsic ellipticities depends already at leading order on both the linear galaxy bias $b$ and the non-linear galaxy bias $b_2$ \citep[e.g.][]{Fry1993}.,
\begin{align}
\label{eq: bispec}
    &\notag B_\mathrm{ggI}(k_1, k_2, k_3; \chi_1, \chi_2, \chi_3) \\
    &= b^2 \, B_{\delta\delta\mathrm{I}}(k_1, k_2, k_3; \chi_1, \chi_2, \chi_3)\\
    &\notag \quad + \frac{b\,b_2}{3}\left[P_{\delta\mathrm{I}}(k_1, \chi_1)\,P(k_2, \chi_2) + P_{\delta\mathrm{I}}(k_1, \chi_1)\,P(k_3, \chi_3)\right. \\
    &\notag \quad \quad \quad \quad \left. + P_{\delta\mathrm{I}}(k_2, \chi_2)\,P(k_3, \chi_3)\right]\\
    &=  b^2\, f_\mathrm{IA}\,  B(k_1, k_2, k_3; \chi_1, \chi_2, \chi_3)\\
    &\notag \quad + f_\mathrm{IA}\,\frac{b\,b_2}{3}\left[P(k_1, \chi_1)\,P(k_2, \chi_2) \right. \\
    &\notag \quad \quad \quad \quad \quad  \left. + P(k_1, \chi_1)\,P(k_3, \chi_3) + P(k_2, \chi_2)\,P(k_3, \chi_3) \right]\;.
\end{align}
This bispectrum contains three free parameters, $A_\mathrm{IA}$, $b$, and $b_2$. However, $A_\mathrm{IA}$ is degenerate with the galaxy bias parameters, as they all determine only the amplitude of the bispectrum. Therefore, we cannot simultaneously constrain the galaxy bias and IA from $\NNMIA$. Instead, we used the value for $b$ estimated by \citet{Singh2015} from second-order statistics. As we argue in Appendix~\ref{app: nonlinear}, we found that for our sample and statistics, the non-linear bias has only a small impact on the model compared to the measurement uncertainty, so we set $b_2$ to zero and kept only $A_\mathrm{IA}$ as a free parameter. We model the matter bispectrum $B$ with the dark-matter-only \verb|bihalofit| prescription \citep{Takahashi2020}. The linear matter power spectrum was modelled using the fitting formula for the transfer function by \citet{Eisenstein1999}. For predicting the LOWZ measurements, we assumed a flat $\Lambda$CDM cosmology with parameters from \citet{Planck2018}, namely $\Omega_\mathrm{m}=0.315$, $\Omega_\mathrm{b}=0.049$, $n_s=0.96$, $\sigma_8=0.811$,  and $h=0.674$.  For predicting the MICE measurements, we used the parameters of the simulation, namely $\Omega_\mathrm{m}=0.25$, $\Omega_\Lambda=0.75$, $\Omega_\mathrm{b}=0.044$, $n_s=0.95$, $\sigma_8=0.8$, $h=0.7$. 

\section{\label{sc:Measurement} Measurement}

\subsection{Correlation function estimator}
We measured $\GGGL^{\Pimax}$ for $N_\mathrm{L}$ lenses and $N_\mathrm{S}$ sources using the estimator 
\begin{align}
\label{eq: biased estimator}
    &\hat{\GGGL}^{\Pimax}_\mathrm{bias}(\rvec_1, \rvec_2) \\
    &\notag= \frac{\sum_{i}^{N_\mathrm{L}} \sum_{j}^{N_\mathrm{L}}  \sum_{k}^{N_\mathrm{S}} \eps_k\,\E^{\I (\phi_i+\phi_j)} \triangle(r_1, r_2, \phi, \Vec{x}_i, \Vec{x}_j, \Vec{x}_k)\,  \Theta_{\Pimax}(\chi_i, \chi_j; \chi_k)}{\sum_{i}^{N_\mathrm{L}} \sum_{j}^{N_\mathrm{L}}  \sum_{k}^{N_\mathrm{S}} \triangle(r_1, r_2, \phi, \Vec{x}_i, \Vec{x}_j, \Vec{x}_k)\, \Theta_{\Pimax}(\chi_i, \chi_j; \chi_k)}\;,
\end{align}
where $\epsilon_k$ is the ellipticity of galaxy $k$, $\triangle$ is one if the galaxy positions form a triangle with side lengths $r_1$ and $r_2$ and opening angle $\phi$ and zero otherwise \footnote{In practice, $\triangle$ is one, if the side lengths of the galaxy triplet triangle fall into a bin centred on $r_1$ and $r_2$, and thus it depends on the chosen bin width. For ease of notation, we omit this dependency here.}, and $\Theta_{\Pimax}$ is one if the lens-source distances are less than or equal to $\Pimax$.  We evaluated the estimator using an adapted version of the \verb|G3LconGPU|\footnote{\url{https://github.com/llinke1/G3LConGPU}} code presented in \citet{Linke2020a}. This code uses GPU acceleration to evaluate the triple sum in Eq.~\eqref{eq: biased estimator} by brute force. Our changes compared to the version in \citet{Linke2020a} consist of implementing a maximum comoving distance $\chi_\mathrm{max}$ between lenses and sources and using physical instead of angular separations. We measure the correlation function for $r_1, r_2$ in 20 logarithmic bins between $0.1\, h^{-1}\, \mathrm{Mpc}$ and $100\, h^{-1}\, \mathrm{Mpc}$ and $\phi$ in 10 linear bins between $0$ and $\pi$.

As shown in App.~\ref{app: bias}, this estimator is biased. Its expectation value is
\begin{align}
    \expval{\hat{\GGGL}^{\Pimax}_\mathrm{bias}} \simeq \frac{ \GGGL^{\Pimax}}{ B(\rvec_1, \rvec_2, \Pimax)}\;,
\end{align}
where 
\begin{align}
    &B(\rvec_1, \rvec_2, \Pimax) \\
    \notag &= \int \dd{\chi} \int_{\chi-\Pimax}^{\chi+\Pimax} \dd{\chi_1} \int_{\chi-\Pimax}^{\chi+\Pimax} \dd{\chi_2}  p(\chi_1)\, p(\chi_2)\, p(\chi)\,\\
    &\notag\quad\times \left[1+\xi(|\rvec_1-\rvec_2|, |\chi_1-\chi_2|)\right]\;,
\end{align}
where $\xi(\rvec, \chi)$ is the three-dimensional two-point correlation function of galaxies with projected separation $\rvec$ and separation $\chi$ along the line of sight. We correct for this bias by multiplying $\hat{\GGGL}^{\Pimax}_\mathrm{bias}$ by an estimated $B$. To obtain this, we estimated the correlation function $\xi$ using a Landy-Szalay estimator as implemented in the code \verb|treecorr| \citep{Jarvis2004}. For the MICE data set, we generate random galaxy positions by uniformly distributing right ascension and declination and drawing redshifts from the redshift distribution of the MICE galaxies. For the LOWZ dataset, we used the same random galaxy sample as \citet{Singh2016}, provided within BOSS DR12. The $\NNMIA$ was estimated by inserting the unbiased estimator $\hat{\GGGL}^{\Pimax}$ into Eq.~\eqref{eq: NNMIA from G}.

\subsection{Covariance estimate}

We estimated the covariance of $\NNMIA$ directly from the data using jackknife resampling, both for the observed LOWZ and simulated MICE galaxies (see Sect.~\ref{sc:Data}). For this, the survey was divided into 100 tiles with approximately the same area. We estimated $\hat{\GGGL}^{\Pimax}$ for each of these tiles individually. We then combine the estimates for all but the $k$-th tile to form the $k$-th jackknife sample and converted it to the aperture statistics, which gave us 100 jackknife samples, we write as $\NNMIA_k$. The $i$-$j$ component of the covariance of $\NNMIA$ is then 
\begin{align}
	\label{eq:covariancematrix}
	\hat{C}_{ij}&= \frac{100}{100-1}\, \sum_{k=1}^{100}\left[\NNMIA_k(R_i)-\overline{\NNMIA_k}(R_i)\right]\,\\
 &\notag \quad \times \left[\NNMIA_k(R_j)-\overline{\NNMIA_k}(R_j)\right]\;,
\end{align}
where $\overline{\NNMIA_k}(\theta_i)$ is the average of all aperture statistics jackknife samples. We used $\sigma_i=\sqrt{\hat{C}
_{ii}}$  as statistical uncertainty on the measured aperture statistics. To estimate the inverse covariance, we applied the Hartlap-correction \citep{Hartlap2007}, i.e., the estimate of the inverse covariance is the inverse of $\hat{C}$ multiplied by the factor
\begin{equation}
    \alpha = \frac{N-p-2}{N-1}\;,
\end{equation}
where $N=100$ is the number of samples and $p$ is the number of entries in the data vector.

\begin{figure}
    \centering
    \includegraphics[width=\linewidth, trim={0.5cm, 0.9cm, 0.5cm, 0.5cm}, clip]{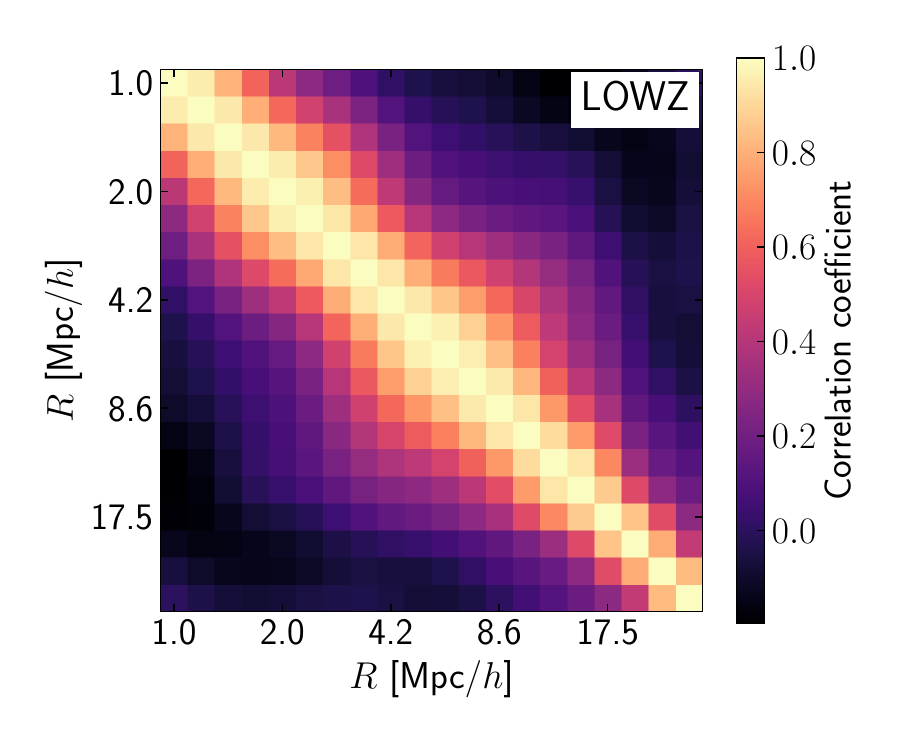}
    \caption{Correlation matrix estimate from jackknife resampling for the LOWZ measurement, using the full sample as shape tracers.}
    \label{fig: corrMat}
\end{figure}

We show the correlation matrix, defined by $\hat{C}_{ij}/(\sigma_i\,\sigma_j)$, for the LOWZ measurement in Fig.~\ref{fig: corrMat}. The correlation matrix is dominated by the diagonal, suggesting that shape noise, in contrast to sample variance, is the biggest contributor. However, we see strong non-diagonal contributions. In particular, the signal for nearby scales is strongly correlated. This is expected for aperture statistics. They combine information from the correlation functions at different scales, which causes correlations between the aperture statistics at different aperture radii. The correlation matrix for $\NNMIA$ is qualitatively similar to the correlation matrix for cosmic shear dominated $\expval{\Nap\Nap\Map}$, shown in Fig. 10 of \citet{Linke2022}.

\subsection{Model fit}
\label{sc:measurement:fit}
We fitted the NLA-inspired model defined in Eq.~\eqref{eq: NNMIA as bispec} and Eq.~\eqref{eq: bispec} to our measurements. For a fixed cosmology, this model includes the galaxy bias $b$ and the IA amplitude $A_\mathrm{IA}$ as free parameters, but we fix the value of $b$ from second-order statistic measurements. For the LOWZ galaxies, we use $b=1.77$ determined by \citet{Singh2015} from second-order statistics. For the MICE galaxies, though, we use $b=1.47$, which is 17\% lower. We use this because the two-point galaxy correlation function in the MICE is approximately 30\% lower than in the LOWZ, which suggests this lower galaxy bias value \citep{Hoffmann2022}. 

Alternatively, one could also infer the galaxy bias and IA amplitude simultaneously from a third-order only analysis. For this, we would require additional measurements of either the position-shape-shape or position-position-position correlations. However, the position-shape-shape correlation has a low signal-to-noise in our sample and cannot be used for this exercise. We also opted against using the position-position-position correlation, even though it is detectable, since this correlation depends even more strongly on non-linear bias terms than the position-position-shape correlation: the relevant bispectrum includes terms proportional to $b^2\,b_2$, not only $b\,b$. Consequently, our assumption of vanishing non-linear bias terms is likely inappropriate for this statistic, and we would need to include $b_2$ as an additional parameter in the model. Additionally, constraints on $b$ from third-order statistics are weaker than those from second-order statistics. Therefore, we can constrain $A_\mathrm{IA}$ to a higher precision when using the $b$ from second-order statistics. Consequently, we decided to use the simpler galaxy bias model and included the $b$ from the second-order measurements. 

Thus, the only free parameter remaining is $A_\mathrm{IA}$, which we obtained by fitting the model to the measurements with a least-squares minimizer \citep{Nelder1965}, including the covariance matrix obtained in Eq.~\eqref{eq:covariancematrix}. To investigate the model validity across different scales, we performed two fits for each measurement. One fit considers the aperture statistics for all scales $R\in [1\,h^{-1}\,\mathrm{Mpc}, 20\,h^{-1}\,\mathrm{Mpc}]$. This includes much smaller scales than the NLA model is considered valid for second-order statistics. We further conducted a second fit, where we limited the measurements to $R>6\,h^{-1}\,\mathrm{Mpc}$. This is the smallest scale considered by \citet{Singh2015} to be valid for the NLA model for second-order statistics. However, we note that the aperture radii $R$ cannot be directly compared to the galaxy separation scale of the two-point correlation function, considered by \citet{Singh2015}. Instead, aperture statistics for radius $R$ include the correlation function both at smaller scales and (depending on the filter function) slightly larger scales than $R$. 

\section{\label{sc:Data} Data}
\subsection{SDSS III BOSS LOWZ galaxies}
Our primary dataset is the SDSS BOSS LOWZ galaxy sample, previously used by \citet{Singh2016}. This sample consists of luminous red galaxies (LRGs) at low redshifts, which were observed by DR8 of the SDSS \citep{Aihara2011} and the BOSS DR11 \citep{Alam2015}. Shape measurements for these galaxies were carried out by \citet{Reyes2012}. They reported galaxy ellipticities $\chi=(1-q^2)/(1+q^2)$, where $q$ is the minor-to-major axis ratio $q$. However, the ellipticity in Eq.~\ref{eq: ellipticity} is defined as $\epsilon=(1-q)/(1+q)$, so we converted $\chi$ to $\epsilon$ using the same relation as \citet{Singh2016},
\begin{equation}
    \epsilon=\frac{\chi}{2\mathcal{R}}\;,
\end{equation}
where $\mathcal{R}=0.925$.

The sample is constructed to contain a volume-limited sample of LRGs, so a selection in both redshift and colour-magnitude space was performed. The selection cuts were
\begin{align}
\label{eq: selection cuts}
    &m_r<13.5+c_\parallel/0.3+\Delta m_r  & 16.0 < m_r < 19.6+\Delta m_r \\
    &\notag |c_\perp| < 0.2    & 0.16 < z < 0.36 \,,
\end{align}
with
\begin{align}
    &c_\parallel =0.7\, (m_g-m_r) + 1.2\, [(m_r-m_i)-0.18] \\
    &\notag c_\perp = (m_r-m_i) - 0.25\, (m_g-m_r) -0.18\;.
\end{align}
For the LOWZ galaxies, $\Delta m_r=0$; however, for the simulated galaxies (see next section), $\Delta m_r$ needs to be set to $0.085$ to match the number density of the observed galaxies. We refer to \citet{Singh2015} for more details on the selection.

Following \citet{Singh2015}, we further divided the galaxy sample into four luminosity bins $L_1$ -- $L_4$ with $L_1$ containing the brightest galaxies. The bins $L_1$ -- $L_3$ each contain 20\% of the galaxies, with $L_4$ containing the remaining 40\% faintest galaxies. In the following we consider correlations between the shapes of galaxies from one of these luminosity bins with the positions of the whole galaxy sample.

\subsection{MICE with IA}
To test our measurement pipeline, we used a catalogue of realistically simulated galaxies based on the MICE grand challenge light cone simulation \citep{Fosalba2015a}. MICE is a dark-matter only $N$-body simulation that adopts a flat $\Lambda$CDM cosmology with parameters $\Omega_\mathrm{m}=0.25$, $\Omega_\Lambda=0.75$, $\Omega_\mathrm{b}=0.044$, $n_s=0.95$, $\sigma_8=0.8$, $h=0.7$. The simulation tracks the evolution of $4096^3$ particles, each with a mass of $2.93 \times 10^{10} h^{-1} M_\odot$, within a cubic volume of side length $3072 h^{-1}$ Mpc, spanning from an initial redshift of $z=100$ to the present time.  Halos were identified with a Friends-of-friends halo finder \citep{Crocce2015}. Subsequently, the halos were populated with galaxies up to redshift $z=1.4$, using a combination of halo abundance matching and a halo occupation distribution model \citep{Carretero2015}. The galaxies received positions, luminosities and colours such that their luminosity function and colour-magnitude distribution matched SDSS observations \citep{Blanton2003, Zehavi2011}.

The MICE was initially designed to accompany gravitational lensing surveys, so an estimate of each galaxy's weak lensing shear was computed by projecting the mass distribution and applying the Born approximation as described in \citet{Fosalba2015b, Fosalba2008}. \citet{Hoffmann2022} added realistic intrinsic galaxy ellipticites using a semi-analytic IA model. In this procedure, the simulated galaxies are divided into red and blue central galaxies and satellites; red centrals receive a 3D orientation aligned with their host halo, while blue centrals are aligned with the angular momentum of their host halo. Satellites are assumed to point towards their host halo centre. These orientations are then distorted by a randomly assigned misalignment angle. The distribution of misalignment angles depends on galaxy colour and magnitude. It is calibrated such that the simulated galaxies reproduce the second-order correlation between intrinsic shapes and galaxy positions observed in the LOWZ galaxy sample as well as the observed distribution of galaxy axis ratios from COSMOS \citep{Laigle2016, Griffith2012}. Simulated galaxy positions, shapes, and other properties for an octant on the sky are available via Cosmohub \footnote{\url{www.cosmohub.pic.es}} \citep{Tallada2020, Carretero2018}

From the MICE with IA, we selected a sample mimicking the LOWZ galaxies on the provided sky octant ($5156.6\,\mathrm{deg}^2$, so approximately half the BOSS footprint). We applied the same selection as in Eq.~\eqref{eq: selection cuts}. However, similar to \citet{Hoffmann2022}, we set $\Delta m_r$ to $0.085$ to match the number density of the LOWZ galaxies. As with the LOWZ sample, we divided the simulated galaxies into four luminosity bins, containing 20\%, 20\%, 20\% and 40\% of the galaxy sample, respectively. We used only the intrinsic galaxy shapes and neglected the cosmic shear contribution. We also used the `true' (or purely cosmological) redshifts of the sample; that is, we neglected any impact of redshift space distortions.

\section{\label{sc:Results} Results}

\begin{figure}
    \centering
    \includegraphics[width=\linewidth, trim={0.7cm 0.75cm 0.7cm 0.5cm}, clip]{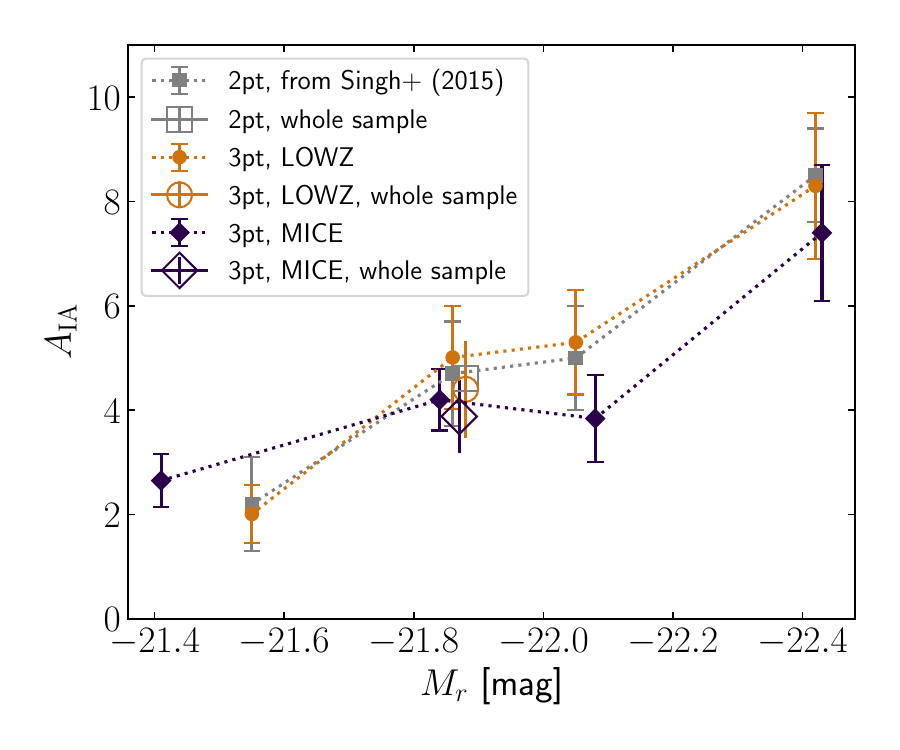}
    \caption{IA amplitude $A_\mathrm{IA}$ as a function of the mean absolute magnitude $M_r$ of the shape tracers for the LOWZ galaxies (pink circles) and for the MICE (green diamonds) as obtained from the third-order IA measurement at scales $R>6\,h^{-1}\,\mathrm{Mpc}$. Also shown are the estimates by \citet{Singh2015} from second-order statistics (black squares). Open symbols are the estimates for the whole shape tracing sample.}
    \label{fig: MR dependence}
\end{figure}

As mentioned in Sect.~\ref{sc:measurement:fit}, we fitted the NLA-inspired model to our measurements firstly using all aperture radii and then restricted to scales above $6\,h^{-1}\,\mathrm{Mpc}$, with galaxy bias values obtained by \citet{Singh2015}. The resulting values for $A_\mathrm{IA}$ are listed in Table~\ref{tab: fit results} and the reduced $\chi^2$ values in Table~\ref{tab: reduced chi2}. In Fig.~\ref{fig: MR dependence}, we show the $A_\mathrm{IA}$ for the larger-scale fits as a function of the mean absolute magnitude $M_r$ of the samples.

\begin{table*}
\caption{Intrinsic alignment amplitudes from model fits to LOWZ and MICE galaxies}
\label{tab: fit results}
\begin{tabular}{l|c|cccc|cccc}
\hline
    & \multicolumn{1}{l|}{} & \multicolumn{4}{c|}{LOWZ}                                                                                      & \multicolumn{4}{c}{MICE}                                                                                       \\
    & $A_\mathrm{IA}$ (2pt) & $\langle M_r \rangle$ & $\langle z \rangle$ & $A_\mathrm{IA}$ (3pt, large scales) & $A_\mathrm{IA}$ (3pt, all) & $\langle M_r \rangle$ & $\langle z \rangle$ & $A_\mathrm{IA}$ (3pt, large scales) & $A_\mathrm{IA}$ (3pt, all) \\ \hline
all & $4.6\pm0.5$          & $-21.88$              & 0.28                & $4.40\pm0.93$                      & $3.99\pm0.53$             & $-21.84$              & 0.26                & $4.20\pm0.59$                      & $3.90\pm0.56$             \\
L1  & $8.5\pm0.9$          & $-22.42$              & 0.28                & $8.3\pm1.4$                        & $\mathbf{7.02\pm0.76}$             & $-22.43$              & 0.29                & $7.4\pm1.3$                        & $\mathbf{6.65\pm0.72}$             \\
L2  & $5.0\pm1.0$          & $-22.05$              & 0.28                & $5.3\pm1.0$                        & $4.57\pm0.58$             & $-22.08$              & 0.28                & $3.84\pm0.83$                      & $4.39\pm0.48$             \\
L3  & $4.7\pm1.0$          & $-21.86$              & 0.28                & $5.01\pm0.99$                      & $\mathbf{3.58\pm0.45}$             & $-21.87$              & 0.27                & $3.88\pm0.70$                      & $4.77\pm0.46$             \\
L4  & $2.2\pm0.9$          & $-21.55$              & 0.28                & $2.01\pm0.56$                      & $1.88\pm0.31$             & $-21.41$              & 0.24                & $3.05\pm0.51$                      & $2.33\pm0.31$             \\ \hline
\end{tabular}
\tablefoot{The $A_\mathrm{IA}$ were obtained by \citet{Singh2015} from second-order statistics (second column), the third-order measurements at $R>6\,h^{-1}\,\mathrm{Mpc}$ (fifth and ninth column),  and at all scales (sixth and tenth column). Also shown are the average absolute magnitude $\langle M_r\rangle$ and the average redshift $\langle z \rangle$ of the samples. Bold denotes values that differ more than 1$\sigma$ from the second-order estimates.}
\end{table*}

For all luminosity bins, $A_\mathrm{IA}$ is significantly non-zero. Using the large-scale (all-scale) fit, $A_\mathrm{IA}$ for the full LOWZ sample is $4.7\sigma$ ($7.6\sigma$) larger than zero. The strongest detection occurs for the brightest sample with $5.9\sigma$ for the large-scale fit. As expected, $A_\mathrm{IA}$ is largest for the brightest galaxies, which are expected to show the strongest alignment, and decreases for fainter galaxies. 

Also shown in Fig.~\ref{fig: MR dependence}, and listed in Table~\ref{tab: fit results}, are the values for $A_\mathrm{IA}$ estimated by \citet{Singh2015} from second-order correlations. Our third-order based estimates are consistent with these second-order based estimates within their $1\sigma$ uncertainties for all luminosity bins. As we are considering the same sample for the second and third-order measurements, the sample variance contributions to the uncertainties of the two cases are correlated. Therefore, we expect a deviation of less than the $1\sigma$ uncertainties. The other contributions to the uncertainties, namely the source galaxies' shape noise and the shot noise of the lens galaxy distribution, contribute differently to the second and third-order measurements and are thus less correlated.

Figure~\ref{fig: MR dependence} also shows the $A_\mathrm{IA}$ obtained from fitting the model to the measurements for the simulated MICE galaxies. Their $A_\mathrm{IA}$ show the same trends as for the observed galaxies and agree with them within $1$--$2\sigma$. Generally, $A_\mathrm{IA}$ is lower in the simulation than observed, except for the faintest galaxies in the L4 sample. However, the MICE galaxies have slightly different mean luminosities and luminosity distributions, so slight differences are not unexpected.

\begin{figure*}
    \centering
    \includegraphics[width=0.7\linewidth]{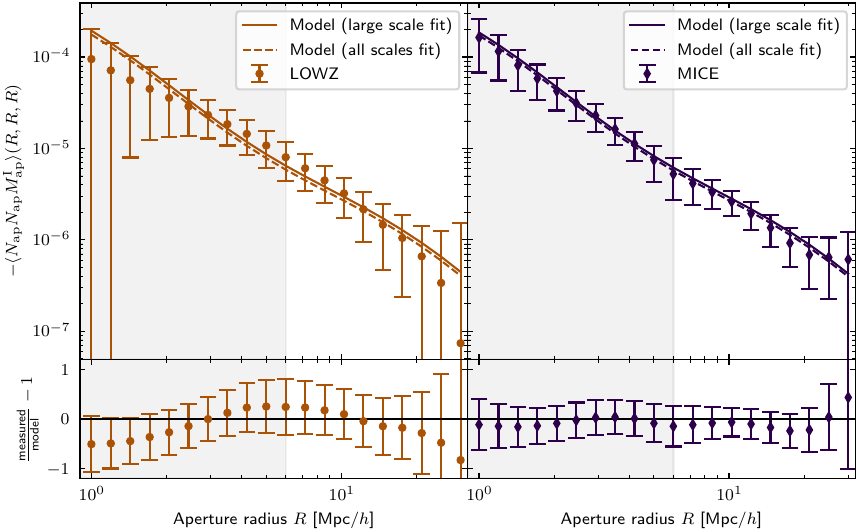}
    \caption{Comparison of IA measurements for full LOWZ and simulated galaxy samples to model fits. Upper panels: IAaperture statistics for full galaxy samples. The left panel shows the measurements for the LOWZ galaxies; the right panel shows measurements for the simulated MICE galaxies. Points are measurements with jackknife error estimates. Solid lines show the model fit using scales above $6\,h^{-1}\,\mathrm{Mpc}$; dashed lines are the model fit using all shown scales. Lower panels: Relative difference of measurement to model fit for large scales.}
    \label{fig: Measurement all}
\end{figure*}

\begin{figure*}
    \centering
    \includegraphics[width=\linewidth]{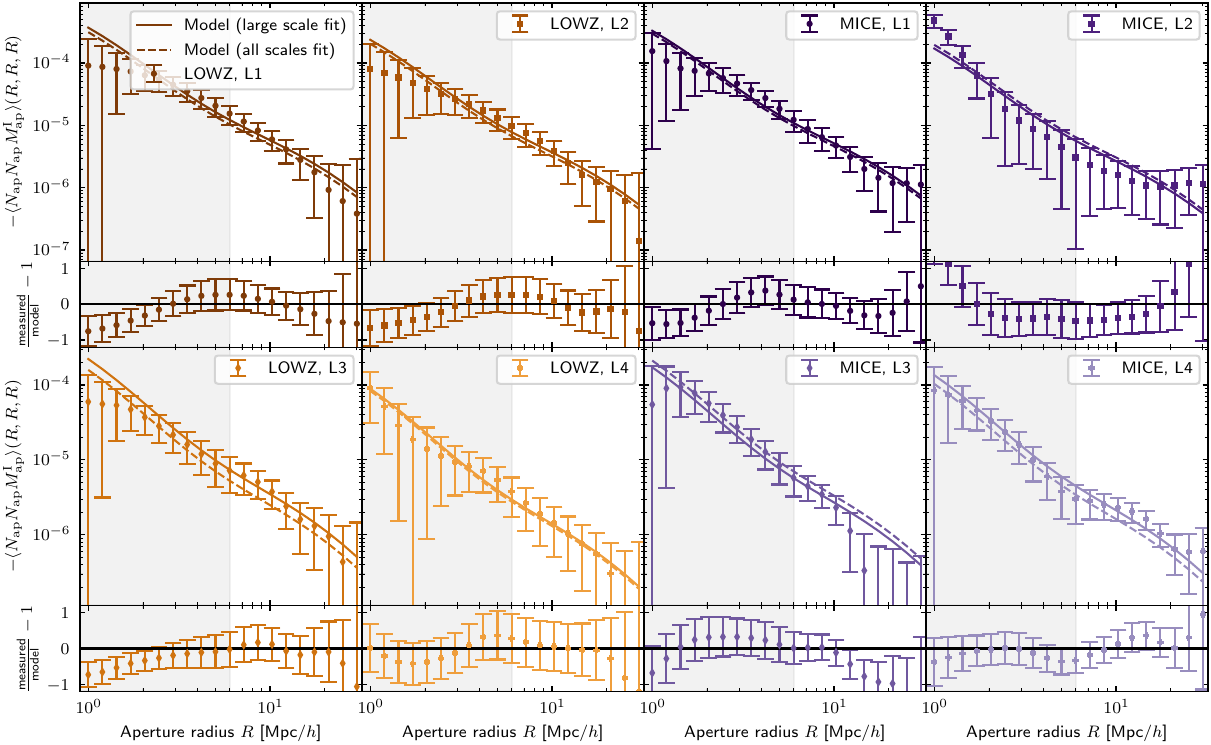}
    \caption{Same as Fig.~\ref{fig: Measurement all}, but for the different luminosity samples, with L1 containing the brightest and L4 the faintest galaxies.}
    \label{fig: Measurement LumBins}
\end{figure*}

Table~\ref{tab: fit results} shows that we obtain smaller values for $A_\mathrm{IA}$ when fitting to all measured scales than when considering only larger scales. To understand this, we show the measurement and both fits for the full sample in Fig.~\ref{fig: Measurement all} and the individual luminosity bins in Fig.~\ref{fig: Measurement LumBins}. For all LOWZ galaxy samples the signal flattens at small scales. This flattening cannot be described by our simple model that only contains a scale-independent parameter. The fit compensates for this inadequacy by lowering $A_\mathrm{IA}$ across all scales.

As can be seen in Table~\ref{tab: reduced chi2}, including the small scales leads to overall worse fits with larger $\chi^2$ values. This confirms that the model cannot accurately describe the small scales. Moreover, the difference between the inferred $A_\mathrm{IA}$ and the estimate from second-order statistics is greater when including the small scales. For the L1 and L3 galaxies, the estimates no longer agree within the $1\sigma$ uncertainty, so the extrapolation of the third-order model to small scales is not consistent with the second-order model at larger scales.

\begin{table}
\centering
\caption{Reduced $\chi^2$ values for model fits}
\label{tab: reduced chi2}
\begin{tabular}{c|cc|cc}
\hline
$\chi^2_\mathrm{reduced}$  & \multicolumn{2}{c|}{LOWZ} & \multicolumn{2}{c}{MICE} \\
                          & large scales    & all     & large scales   & all     \\ \hline
all                       & 0.99           & 1.56   & 0.68          & 1.19   \\
L1                        & 1.19           & 4.20   & 0.93          & 3.83   \\
L2                        & 0.97           & 2.53   & 2.54          & 9.70   \\
L3                        & 0.75           & 2.49   & 2.29          & 5.53   \\
L4                        & 0.84           & 0.93   & 1.29          & 1.35   \\ \hline
\end{tabular}
\end{table}

The MICE galaxies show the same trends as the LOWZ galaxies. Aside from the L2 sample, we see the same flattening of the signal at small scales. For the brightest galaxies, we also find a more than $1\sigma$ difference in the $A_\mathrm{IA}$ inferred from the fit to the whole scale range compared to the second-order estimate.

\section{\label{sc:Discussion} Discussion}
We presented the first detection of a third-order intrinsic alignment (IA) correlation in the Sloan Digital Sky Survey (SDSS) Baryon Oscillation Spectroscopic Survey (BOSS) LOWZ galaxies. For this, we measured the third-order correlation function between galaxy positions and shapes and expressed them in terms of aperture statistics. We compared the measured signal with predictions by the Marenostrum Institut de Ciències de l'Espai (MICE) cosmological simulation and a non-linear linear alignment (NLA) based analytical model. Our measurements depend on the $B_{\delta\delta\mathrm{I}}$ contribution to the bispectrum of matter densities and galaxy shapes, the matter-matter-shape correlation. The positions of the LOWZ galaxies are used as tracers for the matter distribution, which were then correlated with their shapes.

We find a significant third-order IA signal. For scales between $6\,h^{-1}\,\mathrm{Mpc}$ and $20\,h^{-1}\,\mathrm{Mpc}$ and using the full LOWZ galaxy sample, we find an IA amplitude $A_\mathrm{IA}=4.4\pm0.93$, which is a non-zero signal at $4.7\sigma$. After dividing the sample into subsamples based on galaxy luminosity, we find an increased S/N for the brightest galaxies while the S/N for the fainter galaxies decreases. However, even for the faintest galaxies, $A_\mathrm{IA}$ is still non-zero at $3.6\sigma$.  This dependence of the signal on the galaxy luminosity is not surprising, as IA has been shown to increase with galaxy luminosity and stellar mass, both in observations \citep{Singh2015, Singh2016, Johnston2019} and hydrodynamical simulations \citep{Samuroff2021}.

Comparing the measurements from the observations to the predictions by the MICE simulation, we again find that the best-fitting $A_\mathrm{IA}$ increases with luminosity. Indeed, the $A_\mathrm{IA}$s from the simulation agree with the LOWZ measurements at $1$--$2\sigma$. At first glance, this agreement might seem unsurprising as the intrinsic shapes of the simulated galaxies have been tuned to reproduce the second-order IA measurements by \citet{Singh2015} in the LOWZ sample. Moreover, the third-order galaxy clustering of MICE galaxies has been successfully validated \citep{Hoffmann2015}. However, the agreement is a crucial consistency check for two reasons. First, the agreement suggests that tuning the simulation to the second-order statistics automatically provides the correct third-order statistics, lending credibility to the physical model implemented in the simulation. Second, only the intrinsic galaxy shapes were used for the simulation measurements without adding cosmic shear. Consequently, the agreement of the measurements in the observations, where we cannot turn off cosmic shear, shows that weak lensing does not significantly bias our measurement. This validates our choice of $\Pimax$, the maximal distance between galaxies in a triplet considered for the correlation function measurement.  Evidently, the chosen $\Pimax$ is small enough that cosmic shear correlations become sub-dominant. 

Finally, we compared the $A_\mathrm{IA}$ from the model fits to the third-order IA signal to the estimates by \citet{Singh2015} from second-order statistics. The $A_\mathrm{IA}$ obtained from fitting to scales above $6\,h^{-1}\,\mathrm{Mpc}$ agree within $1\sigma$ with the second-order estimates. Consequently, at these scales, the second and third-order models are consistent and can be used in a combined analysis. This combination offers the opportunity for IA self-calibration \citep{Pyne2021}, leading to tighter constraints on both IA and cosmological parameters.

The $A_\mathrm{IA}$ values inferred from the fit, including smaller scales, agree less well with the second-order based estimates. While the agreement is still within $2\sigma$ for all luminosity samples, the measurements show a flattening at small scales that cannot be described with our simple NLA-inspired model. Furthermore, the goodness-of-fit, as indicated by the reduced $\chi^2$ values, degrades when small scales are included. This shows that the model is no longer accurate at small scales. There are several possible reasons for this inaccuracy, among them the inaccuracy of the NLA approach, non-linear contributions to the galaxy bias, and small-scale physical effects such as baryonic feedback. Regarding the NLA model, \citet{Singh2015} find that the NLA model cannot well describe the second-order IA statistics for scales below $6\,h^{-1}\,\mathrm{Mpc}$. Thus, it seems natural that the NLA-inspired model also breaks down at small scales. As mentioned before, the NLA model is an effective, phenomenological description of IA and is not rigorously physically motivated. Higher-order corrections thus become necessary for small scales.

Another consideration in explaining the model deviations is our choice of a simple linear deterministic galaxy bias. In a perturbation theory approach, the galaxy bispectrum already depends on the non-linear galaxy bias at the leading order. We showed that a non-zero $b_2$, which is the first non-linear bias term, only has a small impact on the model compared to the measurement uncertainties and thus can be neglected for the sample and scales we consider here. Higher order contributions, i.e., $b_3$, $b_4$, and so on, enter at most at the next-to-leading order, so they likely have even less impact on the model. Thus, it seems unlikely that the full `flattening' of the signal can be explained by non-linear galaxy bias terms. However,  galaxy bias depends strongly on the selected sample, so while we could simplify our model by neglecting non-linear galaxy bias, this simplification might not be appropriate for different sample choices. We also neglected the quadratic non-local bias, which depends on the tidal field.

Small-scale physics, such as baryonic feedback, could also be part of the cause of the model breakdown at small scales. Baryonic feedback leads to a suppression of the matter power and bispectrum at small scales. This would also suppress the third-order IA signal for small aperture radii. However, the concrete values for the bispectrum suppression are unclear, with predictions from hydrodynamical simulations ranging between less than 1\% and 20\% at $k=1\,h\,\mathrm{Mpc}^{-1}$, while in some simulations there is even an enhancement of the bispectrum at $k\simeq 3\,h\,\mathrm{Mpc}^{-1}$ \citep[see e.g.][]{Arico2020, Foreman2020}.

The perturbation theory approach itself becomes inaccurate when entering the regime of single dark matter halos, so a more sophisticated third-order IA model might be needed at small scales. For example, the bispectrum $B_{\mathrm{ggI}}$ could be modelled by using a halo model approach for the correlation between galaxies and matter \citep{Linke2022}, multiplied by the NLA prefactor $f_1$. Alternatively, a full halo model for the IA signal might be devised in the spirit of \citet{Fortuna2021}. However, in light of the general good agreement, we see no necessity to use a more complex IA model to describe the LOWZ data set. This might change for other surveys with more constraining power, for example, those expected by \Euclid, LSST, and DESI. Repeating our measurements on a more extensive survey and other galaxy samples can help decide whether the NLA-inspired model is still appropriate for analysing third-order statistics in upcoming stage-IV surveys. Additionally, other third-order statistics, for example, shape-shape-lens correlations or the related bispectra, could be explored to further test the model's consistency.

%
%

\begin{acknowledgements}  
We thank the anonymous referee for their helpful comments.
The authors acknowledge the support of the Lorentz Center and the echoIA project.
LL is supported by the Austrian Science Fund (FWF) [ESP 357-N]. 

SP and BJ acknowledge support by the ERC-selected UKRI Frontier Research Grant EP/Y03015X/1.

SP acknowledges  the Lend\"ulet Program of the Hungarian Academy of Sciences and hospitality from
MTA-CSFK Lend\"ulet "Momentum" Large-Scale Structure Research Group, Konkoly Observatory, Budapest.

This publication is part of the project “A rising tide: Galaxy intrinsic alignments as a new probe of cosmology and galaxy evolution'' (with project number VI.Vidi.203.011) of the Talent programme Vidi which is (partly) financed by the Dutch Research Council (NWO).

RM acknowledges the support of the Simons Foundation (Simons Investigator in Astrophysics, Award ID 620789).

This work has made use of CosmoHub. CosmoHub has been developed by the Port d'Informació Científica (PIC), maintained through a collaboration of the Institut de Física d'Altes Energies (IFAE) and the Centro de Investigaciones Energéticas Medioambientales y Tecnológicas (CIEMAT) and the Institute of Space Sciences (CSIC \& IEEC). CosmoHub was partially funded by the "Plan Estatal de Investigación Científica y Técnica de Innovación" program of the Spanish government, has been supported by the call for grants for Scientific and Technical Equipment 2021 of the State Program for Knowledge Generation and Scientific and Technological Strengthening of the R+D+i System, financed by MCIN/AEI/ 10.13039/501100011033 and the EU NextGeneration/PRTR (Hadoop Cluster for the comprehensive management of massive scientific data, reference EQC2021-007479-P) and by MICIIN with funding from European Union NextGenerationEU(PRTR-C17.I1) and by Generalitat de Catalunya.

Funding for SDSS-III has been provided by the Alfred P. Sloan Foundation, the Participating Institutions, the National Science Foundation, and the U.S. Department of Energy Office of Science. The SDSS-III website is http://www.sdss3.org/. SDSS-III is managed by the Astrophysical Research Consortium for the Participating Institutions of the SDSS-III Collaboration including the University of Arizona, the Brazilian Participation Group, Brookhaven National Laboratory, Carnegie Mellon University, University of Florida, the French Participation Group, the German Participation Group, Harvard University, the Instituto de Astrofisica de Canarias, the Michigan State/Notre Dame/JINA Participation Group, Johns Hopkins University, Lawrence Berkeley National Laboratory, Max Planck Institute for Astrophysics, Max Planck Institute for Extraterrestrial Physics, New Mexico State University, New York University, Ohio State University, Pennsylvania State University, University of Portsmouth, Princeton University, the Spanish Participation Group, University of Tokyo, University of Utah, Vanderbilt University, University of Virginia, University of Washington, and Yale University.

\end{acknowledgements}

\bibliography{Euclid}

\begin{thebibliography}{87}
\expandafter\ifx\csname natexlab\endcsname\relax\def\natexlab#1{#1}\fi

\bibitem[{{Abbott} {et~al.}(2016){Abbott}, {Abdalla}, {Allam}, {Amara},
  {Annis}, {Armstrong}, {Bacon}, {Banerji}, {Bauer}, {Baxter}, {Becker},
  {Benoit-L{\'e}vy}, {Bernstein}, {Bernstein}, {Bertin}, {Blazek}, {Bonnett},
  {Bridle}, {Brooks}, {Bruderer}, {Buckley-Geer}, {Burke}, {Busha}, {Capozzi},
  {Carnero Rosell}, {Carrasco Kind}, {Carretero}, {Castander}, {Chang},
  {Clampitt}, {Crocce}, {Cunha}, {D'Andrea}, {da Costa}, {Das}, {DePoy},
  {Desai}, {Diehl}, {Dietrich}, {Dodelson}, {Doel}, {Drlica-Wagner},
  {Efstathiou}, {Eifler}, {Erickson}, {Estrada}, {Evrard}, {Fausti Neto},
  {Fernandez}, {Finley}, {Flaugher}, {Fosalba}, {Friedrich}, {Frieman},
  {Gangkofner}, {Garcia-Bellido}, {Gaztanaga}, {Gerdes}, {Gruen}, {Gruendl},
  {Gutierrez}, {Hartley}, {Hirsch}, {Honscheid}, {Huff}, {Jain}, {James},
  {Jarvis}, {Kacprzak}, {Kent}, {Kirk}, {Krause}, {Kravtsov}, {Kuehn},
  {Kuropatkin}, {Kwan}, {Lahav}, {Leistedt}, {Li}, {Lima}, {Lin}, {MacCrann},
  {March}, {Marshall}, {Martini}, {McMahon}, {Melchior}, {Miller}, {Miquel},
  {Mohr}, {Neilsen}, {Nichol}, {Nicola}, {Nord}, {Ogando}, {Palmese}, {Peiris},
  {Plazas}, {Refregier}, {Roe}, {Romer}, {Roodman}, {Rowe}, {Rykoff}, {Sabiu},
  {Sadeh}, {Sako}, {Samuroff}, {Sanchez}, {S{\'a}nchez}, {Seo},
  {Sevilla-Noarbe}, {Sheldon}, {Smith}, {Soares-Santos}, {Sobreira}, {Suchyta},
  {Swanson}, {Tarle}, {Thaler}, {Thomas}, {Troxel}, {Vikram}, {Walker},
  {Wechsler}, {Weller}, {Zhang}, {Zuntz}, \& {Dark Energy Survey
  Collaboration}}]{Abbott2016}
{Abbott}, T., {Abdalla}, F.~B., {Allam}, S., {et~al.} 2016, \prd, 94, 022001

\bibitem[{{Aihara} {et~al.}(2011){Aihara}, {Allende Prieto}, {An}, {Anderson},
  {Aubourg}, {Balbinot}, {Beers}, {Berlind}, {Bickerton}, {Bizyaev}, {Blanton},
  {Bochanski}, {Bolton}, {Bovy}, {Brandt}, {Brinkmann}, {Brown}, {Brownstein},
  {Busca}, {Campbell}, {Carr}, {Chen}, {Chiappini}, {Comparat}, {Connolly},
  {Cortes}, {Croft}, {Cuesta}, {da Costa}, {Davenport}, {Dawson}, {Dhital},
  {Ealet}, {Ebelke}, {Edmondson}, {Eisenstein}, {Escoffier}, {Esposito},
  {Evans}, {Fan}, {Femen{\'\i}a Castell{\'a}}, {Font-Ribera}, {Frinchaboy},
  {Ge}, {Gillespie}, {Gilmore}, {Gonz{\'a}lez Hern{\'a}ndez}, {Gott}, {Gould},
  {Grebel}, {Gunn}, {Hamilton}, {Harding}, {Harris}, {Hawley}, {Hearty}, {Ho},
  {Hogg}, {Holtzman}, {Honscheid}, {Inada}, {Ivans}, {Jiang}, {Johnson},
  {Jordan}, {Jordan}, {Kazin}, {Kirkby}, {Klaene}, {Knapp}, {Kneib},
  {Kochanek}, {Koesterke}, {Kollmeier}, {Kron}, {Lampeitl}, {Lang}, {Le Goff},
  {Lee}, {Lin}, {Long}, {Loomis}, {Lucatello}, {Lundgren}, {Lupton}, {Ma},
  {MacDonald}, {Mahadevan}, {Maia}, {Makler}, {Malanushenko}, {Malanushenko},
  {Mandelbaum}, {Maraston}, {Margala}, {Masters}, {McBride}, {McGehee},
  {McGreer}, {M{\'e}nard}, {Miralda-Escud{\'e}}, {Morrison}, {Mullally},
  {Muna}, {Munn}, {Murayama}, {Myers}, {Naugle}, {Neto}, {Nguyen}, {Nichol},
  {O'Connell}, {Ogando}, {Olmstead}, {Oravetz}, {Padmanabhan},
  {Palanque-Delabrouille}, {Pan}, {Pandey}, {P{\^a}ris}, {Percival},
  {Petitjean}, {Pfaffenberger}, {Pforr}, {Phleps}, {Pichon}, {Pieri}, {Prada},
  {Price-Whelan}, {Raddick}, {Ramos}, {Reyl{\'e}}, {Rich}, {Richards}, {Rix},
  {Robin}, {Rocha-Pinto}, {Rockosi}, {Roe}, {Rollinde}, {Ross}, {Ross},
  {Rossetto}, {S{\'a}nchez}, {Sayres}, {Schlegel}, {Schlesinger}, {Schmidt},
  {Schneider}, {Sheldon}, {Shu}, {Simmerer}, {Simmons}, {Sivarani}, {Snedden},
  {Sobeck}, {Steinmetz}, {Strauss}, {Szalay}, {Tanaka}, {Thakar}, {Thomas},
  {Tinker}, {Tofflemire}, {Tojeiro}, {Tremonti}, {Vandenberg}, {Vargas
  Maga{\~n}a}, {Verde}, {Vogt}, {Wake}, {Wang}, {Weaver}, {Weinberg}, {White},
  {White}, {Yanny}, {Yasuda}, {Yeche}, \& {Zehavi}}]{Aihara2011}
{Aihara}, H., {Allende Prieto}, C., {An}, D., {et~al.} 2011, \apjs, 193, 29

\bibitem[{{Aihara} {et~al.}(2018){Aihara}, {Arimoto}, {Armstrong}, {Arnouts},
  {Bahcall}, {Bickerton}, {Bosch}, {Bundy}, {Capak}, {Chan}, {Chiba}, {Coupon},
  {Egami}, {Enoki}, {Finet}, {Fujimori}, {Fujimoto}, {Furusawa}, {Furusawa},
  {Goto}, {Goulding}, {Greco}, {Greene}, {Gunn}, {Hamana}, {Harikane},
  {Hashimoto}, {Hattori}, {Hayashi}, {Hayashi}, {He{\l}miniak}, {Higuchi},
  {Hikage}, {Ho}, {Hsieh}, {Huang}, {Huang}, {Ikeda}, {Imanishi}, {Inoue},
  {Iwasawa}, {Iwata}, {Jaelani}, {Jian}, {Kamata}, {Karoji}, {Kashikawa},
  {Katayama}, {Kawanomoto}, {Kayo}, {Koda}, {Koike}, {Kojima}, {Komiyama},
  {Konno}, {Koshida}, {Koyama}, {Kusakabe}, {Leauthaud}, {Lee}, {Lin}, {Lin},
  {Lupton}, {Mandelbaum}, {Matsuoka}, {Medezinski}, {Mineo}, {Miyama},
  {Miyatake}, {Miyazaki}, {Momose}, {More}, {More}, {Moritani}, {Moriya},
  {Morokuma}, {Mukae}, {Murata}, {Murayama}, {Nagao}, {Nakata}, {Niida},
  {Niikura}, {Nishizawa}, {Obuchi}, {Oguri}, {Oishi}, {Okabe}, {Okamoto},
  {Okura}, {Ono}, {Onodera}, {Onoue}, {Osato}, {Ouchi}, {Price}, {Pyo}, {Sako},
  {Sawicki}, {Shibuya}, {Shimasaku}, {Shimono}, {Shirasaki}, {Silverman},
  {Simet}, {Speagle}, {Spergel}, {Strauss}, {Sugahara}, {Sugiyama}, {Suto},
  {Suyu}, {Suzuki}, {Tait}, {Takada}, {Takata}, {Tamura}, {Tanaka}, {Tanaka},
  {Tanaka}, {Tanaka}, {Terai}, {Terashima}, {Toba}, {Tominaga}, {Toshikawa},
  {Turner}, {Uchida}, {Uchiyama}, {Umetsu}, {Uraguchi}, {Urata}, {Usuda},
  {Utsumi}, {Wang}, {Wang}, {Wong}, {Yabe}, {Yamada}, {Yamanoi}, {Yasuda},
  {Yeh}, {Yonehara}, \& {Yuma}}]{Aihara2018}
{Aihara}, H., {Arimoto}, N., {Armstrong}, R., {et~al.} 2018, \pasj, 70, S4

\bibitem[{{Akeson} {et~al.}(2019){Akeson}, {Armus}, {Bachelet}, {Bailey},
  {Bartusek}, {Bellini}, {Benford}, {Bennett}, {Bhattacharya}, {Bohlin},
  {Boyer}, {Bozza}, {Bryden}, {Calchi Novati}, {Carpenter}, {Casertano},
  {Choi}, {Content}, {Dayal}, {Dressler}, {Dor{\'e}}, {Fall}, {Fan}, {Fang},
  {Filippenko}, {Finkelstein}, {Foley}, {Furlanetto}, {Kalirai}, {Gaudi},
  {Gilbert}, {Girard}, {Grady}, {Greene}, {Guhathakurta}, {Heinrich},
  {Hemmati}, {Hendel}, {Henderson}, {Henning}, {Hirata}, {Ho}, {Huff},
  {Hutter}, {Jansen}, {Jha}, {Johnson}, {Jones}, {Kasdin}, {Kelly}, {Kirshner},
  {Koekemoer}, {Kruk}, {Lewis}, {Macintosh}, {Madau}, {Malhotra}, {Mandel},
  {Massara}, {Masters}, {McEnery}, {McQuinn}, {Melchior}, {Melton},
  {Mennesson}, {Peeples}, {Penny}, {Perlmutter}, {Pisani}, {Plazas}, {Poleski},
  {Postman}, {Ranc}, {Rauscher}, {Rest}, {Roberge}, {Robertson}, {Rodney},
  {Rhoads}, {Rhodes}, {Ryan}, {Sahu}, {Sand}, {Scolnic}, {Seth}, {Shvartzvald},
  {Siellez}, {Smith}, {Spergel}, {Stassun}, {Street}, {Strolger}, {Szalay},
  {Trauger}, {Troxel}, {Turnbull}, {van der Marel}, {von der Linden}, {Wang},
  {Weinberg}, {Williams}, {Windhorst}, {Wollack}, {Wu}, {Yee}, \&
  {Zimmerman}}]{Akeson2019}
{Akeson}, R., {Armus}, L., {Bachelet}, E., {et~al.} 2019, arXiv e-prints,
  arXiv:1902.05569

\bibitem[{{Alam} {et~al.}(2015){Alam}, {Albareti}, {Allende Prieto}, {Anders},
  {Anderson}, {Anderton}, {Andrews}, {Armengaud}, {Aubourg}, {Bailey}, {Basu},
  {Bautista}, {Beaton}, {Beers}, {Bender}, {Berlind}, {Beutler}, {Bhardwaj},
  {Bird}, {Bizyaev}, {Blake}, {Blanton}, {Blomqvist}, {Bochanski}, {Bolton},
  {Bovy}, {Shelden Bradley}, {Brandt}, {Brauer}, {Brinkmann}, {Brown},
  {Brownstein}, {Burden}, {Burtin}, {Busca}, {Cai}, {Capozzi}, {Carnero
  Rosell}, {Carr}, {Carrera}, {Chambers}, {Chaplin}, {Chen}, {Chiappini},
  {Chojnowski}, {Chuang}, {Clerc}, {Comparat}, {Covey}, {Croft}, {Cuesta},
  {Cunha}, {da Costa}, {Da Rio}, {Davenport}, {Dawson}, {De Lee}, {Delubac},
  {Deshpande}, {Dhital}, {Dutra-Ferreira}, {Dwelly}, {Ealet}, {Ebelke},
  {Edmondson}, {Eisenstein}, {Ellsworth}, {Elsworth}, {Epstein}, {Eracleous},
  {Escoffier}, {Esposito}, {Evans}, {Fan}, {Fern{\'a}ndez-Alvar}, {Feuillet},
  {Filiz Ak}, {Finley}, {Finoguenov}, {Flaherty}, {Fleming}, {Font-Ribera},
  {Foster}, {Frinchaboy}, {Galbraith-Frew}, {Garc{\'\i}a},
  {Garc{\'\i}a-Hern{\'a}ndez}, {Garc{\'\i}a P{\'e}rez}, {Gaulme}, {Ge},
  {G{\'e}nova-Santos}, {Georgakakis}, {Ghezzi}, {Gillespie}, {Girardi},
  {Goddard}, {Gontcho}, {Gonz{\'a}lez Hern{\'a}ndez}, {Grebel}, {Green},
  {Grieb}, {Grieves}, {Gunn}, {Guo}, {Harding}, {Hasselquist}, {Hawley},
  {Hayden}, {Hearty}, {Hekker}, {Ho}, {Hogg}, {Holley-Bockelmann}, {Holtzman},
  {Honscheid}, {Huber}, {Huehnerhoff}, {Ivans}, {Jiang}, {Johnson},
  {Kinemuchi}, {Kirkby}, {Kitaura}, {Klaene}, {Knapp}, {Kneib}, {Koenig},
  {Lam}, {Lan}, {Lang}, {Laurent}, {Le Goff}, {Leauthaud}, {Lee}, {Lee},
  {Licquia}, {Liu}, {Long}, {L{\'o}pez-Corredoira}, {Lorenzo-Oliveira},
  {Lucatello}, {Lundgren}, {Lupton}, {Mack}, {Mahadevan}, {Maia}, {Majewski},
  {Malanushenko}, {Malanushenko}, {Manchado}, {Manera}, {Mao}, {Maraston},
  {Marchwinski}, {Margala}, {Martell}, {Martig}, {Masters}, {Mathur},
  {McBride}, {McGehee}, {McGreer}, {McMahon}, {M{\'e}nard}, {Menzel},
  {Merloni}, {M{\'e}sz{\'a}ros}, {Miller}, {Miralda-Escud{\'e}}, {Miyatake},
  {Montero-Dorta}, {More}, {Morganson}, {Morice-Atkinson}, {Morrison},
  {Mosser}, {Muna}, {Myers}, {Nandra}, {Newman}, {Neyrinck}, {Nguyen},
  {Nichol}, {Nidever}, {Noterdaeme}, {Nuza}, {O'Connell}, {O'Connell},
  {O'Connell}, {Ogando}, {Olmstead}, {Oravetz}, {Oravetz}, {Osumi}, {Owen},
  {Padgett}, {Padmanabhan}, {Paegert}, {Palanque-Delabrouille}, {Pan},
  {Parejko}, {P{\^a}ris}, {Park}, {Pattarakijwanich}, {Pellejero-Ibanez},
  {Pepper}, {Percival}, {P{\'e}rez-Fournon}, {P{\'e}rez-R{\`a}fols},
  {Petitjean}, {Pieri}, {Pinsonneault}, {Porto de Mello}, {Prada}, {Prakash},
  {Price-Whelan}, {Protopapas}, {Raddick}, {Rahman}, {Reid}, {Rich}, {Rix},
  {Robin}, {Rockosi}, {Rodrigues}, {Rodr{\'\i}guez-Torres}, {Roe}, {Ross},
  {Ross}, {Rossi}, {Ruan}, {Rubi{\~n}o-Mart{\'\i}n}, {Rykoff},
  {Salazar-Albornoz}, {Salvato}, {Samushia}, {S{\'a}nchez}, {Santiago},
  {Sayres}, {Schiavon}, {Schlegel}, {Schmidt}, {Schneider}, {Schultheis},
  {Schwope}, {Sc{\'o}ccola}, {Scott}, {Sellgren}, {Seo}, {Serenelli}, {Shane},
  {Shen}, {Shetrone}, {Shu}, {Silva Aguirre}, {Sivarani}, {Skrutskie},
  {Slosar}, {Smith}, {Sobreira}, {Souto}, {Stassun}, {Steinmetz}, {Stello},
  {Strauss}, {Streblyanska}, {Suzuki}, {Swanson}, {Tan}, {Tayar}, {Terrien},
  {Thakar}, {Thomas}, {Thomas}, {Thompson}, {Tinker}, {Tojeiro}, {Troup},
  {Vargas-Maga{\~n}a}, {Vazquez}, {Verde}, {Viel}, {Vogt}, {Wake}, {Wang},
  {Weaver}, {Weinberg}, {Weiner}, {White}, {Wilson}, {Wisniewski},
  {Wood-Vasey}, {Ye`che}, {York}, {Zakamska}, {Zamora}, {Zasowski}, {Zehavi},
  {Zhao}, {Zheng}, {Zhou}, {Zhou}, {Zou}, \& {Zhu}}]{Alam2015}
{Alam}, S., {Albareti}, F.~D., {Allende Prieto}, C., {et~al.} 2015, \apjs, 219,
  12

\bibitem[{{Amon} {et~al.}(2022){Amon}, {Gruen}, {Troxel}, {MacCrann},
  {Dodelson}, {Choi}, {Doux}, {Secco}, {Samuroff}, {Krause}, {Cordero},
  {Myles}, {DeRose}, {Wechsler}, {Gatti}, {Navarro-Alsina}, {Bernstein},
  {Jain}, {Blazek}, {Alarcon}, {Fert{\'e}}, {Lemos}, {Raveri}, {Campos},
  {Prat}, {S{\'a}nchez}, {Jarvis}, {Alves}, {Andrade-Oliveira}, {Baxter},
  {Bechtol}, {Becker}, {Bridle}, {Camacho}, {Carnero Rosell}, {Carrasco Kind},
  {Cawthon}, {Chang}, {Chen}, {Chintalapati}, {Crocce}, {Davis}, {Diehl},
  {Drlica-Wagner}, {Eckert}, {Eifler}, {Elvin-Poole}, {Everett}, {Fang},
  {Fosalba}, {Friedrich}, {Gaztanaga}, {Giannini}, {Gruendl}, {Harrison},
  {Hartley}, {Herner}, {Huang}, {Huff}, {Huterer}, {Kuropatkin}, {Leget},
  {Liddle}, {McCullough}, {Muir}, {Pandey}, {Park}, {Porredon}, {Refregier},
  {Rollins}, {Roodman}, {Rosenfeld}, {Ross}, {Rykoff}, {Sanchez},
  {Sevilla-Noarbe}, {Sheldon}, {Shin}, {Troja}, {Tutusaus}, {Tutusaus},
  {Varga}, {Weaverdyck}, {Yanny}, {Yin}, {Zhang}, {Zuntz}, {Aguena}, {Allam},
  {Annis}, {Bacon}, {Bertin}, {Bhargava}, {Brooks}, {Buckley-Geer}, {Burke},
  {Carretero}, {Costanzi}, {da Costa}, {Pereira}, {De Vicente}, {Desai},
  {Dietrich}, {Doel}, {Ferrero}, {Flaugher}, {Frieman}, {Garc{\'\i}a-Bellido},
  {Gaztanaga}, {Gerdes}, {Giannantonio}, {Gschwend}, {Gutierrez}, {Hinton},
  {Hollowood}, {Honscheid}, {Hoyle}, {James}, {Kron}, {Kuehn}, {Lahav}, {Lima},
  {Lin}, {Maia}, {Marshall}, {Martini}, {Melchior}, {Menanteau}, {Miquel},
  {Mohr}, {Morgan}, {Ogando}, {Palmese}, {Paz-Chinch{\'o}n}, {Petravick},
  {Pieres}, {Romer}, {Sanchez}, {Scarpine}, {Schubnell}, {Serrano}, {Smith},
  {Soares-Santos}, {Tarle}, {Thomas}, {To}, {Weller}, \& {DES
  Collaboration}}]{Amon2022}
{Amon}, A., {Gruen}, D., {Troxel}, M.~A., {et~al.} 2022, \prd, 105, 023514

\bibitem[{{Aric{\`o}} {et~al.}(2021){Aric{\`o}}, {Angulo},
  {Hern{\'a}ndez-Monteagudo}, {Contreras}, \& {Zennaro}}]{Arico2020}
{Aric{\`o}}, G., {Angulo}, R.~E., {Hern{\'a}ndez-Monteagudo}, C., {Contreras},
  S., \& {Zennaro}, M. 2021, \mnras, 503, 3596

\bibitem[{{Asgari} {et~al.}(2021){Asgari}, {Lin}, {Joachimi}, {Giblin},
  {Heymans}, {Hildebrandt}, {Kannawadi}, {St{\"o}lzner}, {Tr{\"o}ster}, {van
  den Busch}, {Wright}, {Bilicki}, {Blake}, {de Jong}, {Dvornik}, {Erben},
  {Getman}, {Hoekstra}, {K{\"o}hlinger}, {Kuijken}, {Miller}, {Radovich},
  {Schneider}, {Shan}, \& {Valentijn}}]{Asgari2021}
{Asgari}, M., {Lin}, C.-A., {Joachimi}, B., {et~al.} 2021, \aap, 645, A104

\bibitem[{{Bartelmann} \& {Schneider}(2001)}]{Bartelmann2001}
{Bartelmann}, M. \& {Schneider}, P. 2001, \physrep, 340, 291

\bibitem[{{Becker} {et~al.}(2016){Becker}, {Troxel}, {MacCrann}, {Krause},
  {Eifler}, {Friedrich}, {Nicola}, {Refregier}, {Amara}, {Bacon}, {Bernstein},
  {Bonnett}, {Bridle}, {Busha}, {Chang}, {Dodelson}, {Erickson}, {Evrard},
  {Frieman}, {Gaztanaga}, {Gruen}, {Hartley}, {Jain}, {Jarvis}, {Kacprzak},
  {Kirk}, {Kravtsov}, {Leistedt}, {Peiris}, {Rykoff}, {Sabiu}, {S{\'a}nchez},
  {Seo}, {Sheldon}, {Wechsler}, {Zuntz}, {Abbott}, {Abdalla}, {Allam},
  {Armstrong}, {Banerji}, {Bauer}, {Benoit-L{\'e}vy}, {Bertin}, {Brooks},
  {Buckley-Geer}, {Burke}, {Capozzi}, {Carnero Rosell}, {Carrasco Kind},
  {Carretero}, {Castander}, {Crocce}, {Cunha}, {D'Andrea}, {da Costa}, {DePoy},
  {Desai}, {Diehl}, {Dietrich}, {Doel}, {Fausti Neto}, {Fernandez}, {Finley},
  {Flaugher}, {Fosalba}, {Gerdes}, {Gruendl}, {Gutierrez}, {Honscheid},
  {James}, {Kuehn}, {Kuropatkin}, {Lahav}, {Li}, {Lima}, {Maia}, {March},
  {Martini}, {Melchior}, {Miller}, {Miquel}, {Mohr}, {Nichol}, {Nord},
  {Ogando}, {Plazas}, {Reil}, {Romer}, {Roodman}, {Sako}, {Sanchez},
  {Scarpine}, {Schubnell}, {Sevilla-Noarbe}, {Smith}, {Soares-Santos},
  {Sobreira}, {Suchyta}, {Swanson}, {Tarle}, {Thaler}, {Thomas}, {Vikram},
  {Walker}, \& {Dark Energy Survey Collaboration}}]{Becker2016}
{Becker}, M.~R., {Troxel}, M.~A., {MacCrann}, N., {et~al.} 2016, \prd, 94,
  022002

\bibitem[{{Blanton} {et~al.}(2003){Blanton}, {Hogg}, {Bahcall}, {Brinkmann},
  {Britton}, {Connolly}, {Csabai}, {Fukugita}, {Loveday}, {Meiksin}, {Munn},
  {Nichol}, {Okamura}, {Quinn}, {Schneider}, {Shimasaku}, {Strauss}, {Tegmark},
  {Vogeley}, \& {Weinberg}}]{Blanton2003}
{Blanton}, M.~R., {Hogg}, D.~W., {Bahcall}, N.~A., {et~al.} 2003, \apj, 592,
  819

\bibitem[{{Blazek} {et~al.}(2019){Blazek}, {MacCrann}, {Troxel}, \&
  {Fang}}]{Blazek2019}
{Blazek}, J.~A., {MacCrann}, N., {Troxel}, M.~A., \& {Fang}, X. 2019, \prd,
  100, 103506

\bibitem[{{Bridle} \& {King}(2007)}]{Bridle2007}
{Bridle}, S. \& {King}, L. 2007, New Journal of Physics, 9, 444

\bibitem[{{Burger} {et~al.}(2024){Burger}, {Porth}, {Heydenreich}, {Linke},
  {Wielders}, {Schneider}, {Asgari}, {Castro}, {Dolag}, {Harnois-D{\'e}raps},
  {Hildebrandt}, {Kuijken}, \& {Martinet}}]{Burger2024}
{Burger}, P.~A., {Porth}, L., {Heydenreich}, S., {et~al.} 2024, \aap, 683, A103

\bibitem[{{Carretero} {et~al.}(2015){Carretero}, {Castander}, {Gazta{\~n}aga},
  {Crocce}, \& {Fosalba}}]{Carretero2015}
{Carretero}, J., {Castander}, F.~J., {Gazta{\~n}aga}, E., {Crocce}, M., \&
  {Fosalba}, P. 2015, \mnras, 447, 646

\bibitem[{{Carretero} {et~al.}(2017){Carretero}, {Tallada}, {Casals}, {Caubet},
  {Castander}, {Blot}, {Alarc{\'o}n}, {Serrano}, {Fosalba}, {Acosta-Silva},
  {Tonello}, {Torradeflot}, {Eriksen}, {Neissner}, \&
  {Delfino}}]{Carretero2018}
{Carretero}, J., {Tallada}, P., {Casals}, J., {et~al.} 2017, in Proceedings of
  the European Physical Society Conference on High Energy Physics. 5-12 July,
  488

\bibitem[{{Catelan} {et~al.}(2001){Catelan}, {Kamionkowski}, \&
  {Blandford}}]{Catelan2001}
{Catelan}, P., {Kamionkowski}, M., \& {Blandford}, R.~D. 2001, \mnras, 320, L7

\bibitem[{{Chen} \& {Kokron}(2024)}]{Chen2024}
{Chen}, S.-F. \& {Kokron}, N. 2024, \jcap, 1, 027

\bibitem[{{Crittenden} {et~al.}(2002){Crittenden}, {Natarajan}, {Pen}, \&
  {Theuns}}]{Crittenden2002}
{Crittenden}, R.~G., {Natarajan}, P., {Pen}, U.-L., \& {Theuns}, T. 2002, \apj,
  568, 20

\bibitem[{{Crocce} {et~al.}(2015){Crocce}, {Castander}, {Gazta{\~n}aga},
  {Fosalba}, \& {Carretero}}]{Crocce2015}
{Crocce}, M., {Castander}, F.~J., {Gazta{\~n}aga}, E., {Fosalba}, P., \&
  {Carretero}, J. 2015, \mnras, 453, 1513

\bibitem[{{Dalal} {et~al.}(2023){Dalal}, {Li}, {Nicola}, {Zuntz}, {Strauss},
  {Sugiyama}, {Zhang}, {Rau}, {Mandelbaum}, {Takada}, {More}, {Miyatake},
  {Kannawadi}, {Shirasaki}, {Taniguchi}, {Takahashi}, {Osato}, {Hamana},
  {Oguri}, {Nishizawa}, {Malag{\'o}n}, {Sunayama}, {Alonso}, {Slosar}, {Luo},
  {Armstrong}, {Bosch}, {Hsieh}, {Komiyama}, {Lupton}, {Lust}, {MacArthur},
  {Miyazaki}, {Murayama}, {Nishimichi}, {Okura}, {Price}, {Tait}, {Tanaka}, \&
  {Wang}}]{Dalal2023}
{Dalal}, R., {Li}, X., {Nicola}, A., {et~al.} 2023, \prd, 108, 123519

\bibitem[{{DES and KiDS Collaboration: Abbott} {et~al.}(2023){DES and KiDS
  Collaboration: Abbott}, {Aguena}, {Alarcon}, {Alves}, {Amon},
  {Andrade-Oliveira}, {Asgari}, {Avila}, {Bacon}, {Bechtol}, {Becker},
  {Bernstein}, {Bertin}, {Bilicki}, {Blazek}, {Bocquet}, {Brooks}, {Burger},
  {Burke}, {Camacho}, {Campos}, {Carnero Rosell}, {Carrasco Kind}, {Carretero},
  {Castander}, {Cawthon}, {Chang}, {Chen}, {Choi}, {Conselice}, {Cordero},
  {Crocce}, {da Costa}, {da Silva Pereira}, {Dalal}, {Davis}, {de Jong},
  {DeRose}, {Desai}, {Diehl}, {Dodelson}, {Doel}, {Doux}, {Drlica-Wagner},
  {Dvornik}, {Eckert}, {Eifler}, {Elvin-Poole}, {Everett}, {Fang}, {Ferrero},
  {Fert{\'e}}, {Flaugher}, {Friedrich}, {Frieman}, {Garc{\'\i}a-Bellido},
  {Gatti}, {Giannini}, {Giblin}, {Gruen}, {Gruendl}, {Gutierrez}, {Harrison},
  {Hartley}, {Herner}, {Heymans}, {Hildebrandt}, {Hinton}, {Hoekstra},
  {Hollowood}, {Honscheid}, {Huang}, {Huff}, {Huterer}, {James}, {Jarvis},
  {Jeffrey}, {Jeltema}, {Joachimi}, {Joudaki}, {Kannawadi}, {Krause}, {Kuehn},
  {Kuijken}, {Kuropatkin}, {Lahav}, {Leget}, {Lemos}, {Li}, {Li}, {Liddle},
  {Lima}, {Lin}, {Lin}, {MacCrann}, {Mahony}, {Marshall}, {McCullough},
  {Mena-Fern{\'a}ndez}, {Menanteau}, {Miquel}, {Mohr}, {Muir}, {Myles},
  {Napolitano}, {Navarro-Alsina}, {Ogando}, {Palmese}, {Pandey}, {Park},
  {Paterno}, {Peacock}, {Petravick}, {Pieres}, {Plazas Malag{\'o}n},
  {Porredon}, {Prat}, {Radovich}, {Raveri}, {Reischke}, {Robertson}, {Rollins},
  {Romer}, {Roodman}, {Rykoff}, {Samuroff}, {S{\'a}nchez}, {Sanchez},
  {Sanchez}, {Schneider}, {Secco}, {Sevilla-Noarbe}, {Shan}, {Sheldon}, {Shin},
  {Sif{\'o}n}, {Smith}, {Soares-Santos}, {St{\"o}lzner}, {Suchyta}, {Swanson},
  {Tarle}, {Thomas}, {To}, {Troxel}, {Tr{\"o}ster}, {Tutusaus}, {van den
  Busch}, {Varga}, {Walker}, {Weaverdyck}, {Wechsler}, {Weller}, {Wiseman},
  {Wright}, {Yanny}, {Yin}, {Yoon}, {Zhang}, \& {Zuntz}}]{DES_KiDS2023}
{DES and KiDS Collaboration: Abbott}, T.~M.~C., {Aguena}, M., {Alarcon}, A.,
  {et~al.} 2023, The Open Journal of Astrophysics, 6, 36

\bibitem[{{Eggemeier} {et~al.}(2021){Eggemeier}, {Scoccimarro}, {Smith},
  {Crocce}, {Pezzotta}, \& {S{\'a}nchez}}]{Eggemeier2021}
{Eggemeier}, A., {Scoccimarro}, R., {Smith}, R.~E., {et~al.} 2021, \prd, 103,
  123550

\bibitem[{{Eisenstein} \& {Hu}(1999)}]{Eisenstein1999}
{Eisenstein}, D.~J. \& {Hu}, W. 1999, \apj, 511, 5

\bibitem[{{Euclid Collaboration: Ajani} {et~al.}(2023){Euclid Collaboration:
  Ajani}, {Baldi}, {Barthelemy}, {Boyle}, {Burger}, {Cardone}, {Cheng},
  {Codis}, {Giocoli}, {Harnois-D{\'e}raps}, {Heydenreich}, {Kansal},
  {Kilbinger}, {Linke}, {Llinares}, {Martinet}, {Parroni}, {Peel}, {Pires},
  {Porth}, {Tereno}, {Uhlemann}, {Vicinanza}, {Vinciguerra}, {Aghanim},
  {Auricchio}, {Bonino}, {Branchini}, {Brescia}, {Brinchmann}, {Camera},
  {Capobianco}, {Carbone}, {Carretero}, {Castander}, {Castellano}, {Cavuoti},
  {Cimatti}, {Cledassou}, {Congedo}, {Conselice}, {Conversi}, {Corcione},
  {Courbin}, {Cropper}, {Da Silva}, {Degaudenzi}, {Di Giorgio}, {Dinis},
  {Douspis}, {Dubath}, {Dupac}, {Farrens}, {Ferriol}, {Fosalba}, {Frailis},
  {Franceschi}, {Galeotta}, {Garilli}, {Gillis}, {Grazian}, {Grupp},
  {Hoekstra}, {Holmes}, {Hornstrup}, {Hudelot}, {Jahnke}, {Jhabvala},
  {K{\"u}mmel}, {Kitching}, {Kunz}, {Kurki-Suonio}, {Lilje}, {Lloro},
  {Maiorano}, {Mansutti}, {Marggraf}, {Markovic}, {Marulli}, {Massey}, {Mei},
  {Mellier}, {Meneghetti}, {Moresco}, {Moscardini}, {Niemi}, {Nightingale},
  {Nutma}, {Padilla}, {Paltani}, {Pedersen}, {Pettorino}, {Polenta}, {Poncet},
  {Popa}, {Raison}, {Renzi}, {Rhodes}, {Riccio}, {Romelli}, {Roncarelli},
  {Rossetti}, {Saglia}, {Sapone}, {Sartoris}, {Schneider}, {Schrabback},
  {Secroun}, {Seidel}, {Serrano}, {Sirignano}, {Stanco}, {Starck},
  {Tallada-Cresp{\'\i}}, {Taylor}, {Toledo-Moreo}, {Torradeflot}, {Tutusaus},
  {Valentijn}, {Valenziano}, {Vassallo}, {Wang}, {Weller}, {Zamorani},
  {Zoubian}, {Andreon}, {Bardelli}, {Boucaud}, {Bozzo}, {Colodro-Conde}, {Di
  Ferdinando}, {Fabbian}, {Farina}, {Graci{\'a}-Carpio}, {Keih{\"a}nen},
  {Lindholm}, {Maino}, {Mauri}, {Neissner}, {Schirmer}, {Scottez}, {Zucca},
  {Akrami}, {Baccigalupi}, {Balaguera-Antol{\'\i}nez}, {Ballardini},
  {Bernardeau}, {Biviano}, {Blanchard}, {Borgani}, {Borlaff}, {Burigana},
  {Cabanac}, {Cappi}, {Carvalho}, {Casas}, {Castignani}, {Castro}, {Chambers},
  {Cooray}, {Coupon}, {Courtois}, {Davini}, {de la Torre}, {De Lucia},
  {Desprez}, {Dole}, {Escartin}, {Escoffier}, {Ferrero}, {Finelli}, {Ganga},
  {Garcia-Bellido}, {George}, {Giacomini}, {Gozaliasl}, {Hildebrandt}, {Jimenez
  Mu{\~n}oz}, {Joachimi}, {Kajava}, {Kirkpatrick}, {Legrand}, {Loureiro},
  {Magliocchetti}, {Maoli}, {Marcin}, {Martinelli}, {Martins}, {Matthew},
  {Maurin}, {Metcalf}, {Monaco}, {Morgante}, {Nadathur}, {Nucita}, {Popa},
  {Potter}, {Pourtsidou}, {P{\"o}ntinen}, {Reimberg}, {S{\'a}nchez}, {Sakr},
  {Schneider}, {Sefusatti}, {Sereno}, {Shulevski}, {Spurio Mancini},
  {Steinwagner}, {Teyssier}, {Valiviita}, {Veropalumbo}, {Viel}, \&
  {Zinchenko}}]{Ajani-EP29}
{Euclid Collaboration: Ajani}, V., {Baldi}, M., {Barthelemy}, A., {et~al.}
  2023, \aap, 675, A120

\bibitem[{{Euclid Collaboration: Blanchard} {et~al.}(2020){Euclid
  Collaboration: Blanchard}, {Camera}, {Carbone}, {Cardone}, {Casas}, {Clesse},
  {Ili{\'c}}, {Kilbinger}, {Kitching}, {Kunz}, {Lacasa}, {Linder}, {Majerotto},
  {Markovi{\v{c}}}, {Martinelli}, {Pettorino}, {Pourtsidou}, {Sakr},
  {S{\'a}nchez}, {Sapone}, {Tutusaus}, {Yahia-Cherif}, {Yankelevich},
  {Andreon}, {Aussel}, {Balaguera-Antol{\'\i}nez}, {Baldi}, {Bardelli},
  {Bender}, {Biviano}, {Bonino}, {Boucaud}, {Bozzo}, {Branchini}, {Brau-Nogue},
  {Brescia}, {Brinchmann}, {Burigana}, {Cabanac}, {Capobianco}, {Cappi},
  {Carretero}, {Carvalho}, {Casas}, {Castander}, {Castellano}, {Cavuoti},
  {Cimatti}, {Cledassou}, {Colodro-Conde}, {Congedo}, {Conselice}, {Conversi},
  {Copin}, {Corcione}, {Coupon}, {Courtois}, {Cropper}, {Da Silva}, {de la
  Torre}, {Di Ferdinando}, {Dubath}, {Ducret}, {Duncan}, {Dupac}, {Dusini},
  {Fabbian}, {Fabricius}, {Farrens}, {Fosalba}, {Fotopoulou}, {Fourmanoit},
  {Frailis}, {Franceschi}, {Franzetti}, {Fumana}, {Galeotta}, {Gillard},
  {Gillis}, {Giocoli}, {G{\'o}mez-Alvarez}, {Graci{\'a}-Carpio}, {Grupp},
  {Guzzo}, {Hoekstra}, {Hormuth}, {Israel}, {Jahnke}, {Keihanen}, {Kermiche},
  {Kirkpatrick}, {Kohley}, {Kubik}, {Kurki-Suonio}, {Ligori}, {Lilje}, {Lloro},
  {Maino}, {Maiorano}, {Marggraf}, {Martinet}, {Marulli}, {Massey},
  {Medinaceli}, {Mei}, {Mellier}, {Metcalf}, {Metge}, {Meylan}, {Moresco},
  {Moscardini}, {Munari}, {Nichol}, {Niemi}, {Nucita}, {Padilla}, {Paltani},
  {Pasian}, {Percival}, {Pires}, {Polenta}, {Poncet}, {Pozzetti}, {Racca},
  {Raison}, {Renzi}, {Rhodes}, {Romelli}, {Roncarelli}, {Rossetti}, {Saglia},
  {Schneider}, {Scottez}, {Secroun}, {Sirri}, {Stanco}, {Starck}, {Sureau},
  {Tallada-Cresp{\'\i}}, {Tavagnacco}, {Taylor}, {Tenti}, {Tereno},
  {Toledo-Moreo}, {Torradeflot}, {Valenziano}, {Vassallo}, {Verdoes Kleijn},
  {Viel}, {Wang}, {Zacchei}, {Zoubian}, \& {Zucca}}]{Blanchard-EP7}
{Euclid Collaboration: Blanchard}, A., {Camera}, S., {Carbone}, C., {et~al.}
  2020, \aap, 642, A191

\bibitem[{{Euclid Collaboration: Mellier} {et~al.}(2024){Euclid Collaboration:
  Mellier}, {Abdurro'uf}, {Acevedo~Barroso}, {et~al.}}]{EuclidSkyOverview}
{Euclid Collaboration: Mellier}, Y., {Abdurro'uf}, {Acevedo~Barroso}, J.,
  {et~al.} 2024, \aap, accepted, arXiv:2405.13491

\bibitem[{{Foreman} {et~al.}(2020){Foreman}, {Coulton}, {Villaescusa-Navarro},
  \& {Barreira}}]{Foreman2020}
{Foreman}, S., {Coulton}, W., {Villaescusa-Navarro}, F., \& {Barreira}, A.
  2020, \mnras, 498, 2887

\bibitem[{{Fortuna} {et~al.}(2021){Fortuna}, {Hoekstra}, {Johnston}, {Vakili},
  {Kannawadi}, {Georgiou}, {Joachimi}, {Wright}, {Asgari}, {Bilicki},
  {Heymans}, {Hildebrandt}, {Kuijken}, \& {Von
  Wietersheim-Kramsta}}]{Fortuna2021}
{Fortuna}, M.~C., {Hoekstra}, H., {Johnston}, H., {et~al.} 2021, \aap, 654, A76

\bibitem[{{Fosalba} {et~al.}(2015{\natexlab{a}}){Fosalba}, {Crocce},
  {Gazta{\~n}aga}, \& {Castander}}]{Fosalba2015a}
{Fosalba}, P., {Crocce}, M., {Gazta{\~n}aga}, E., \& {Castander}, F.~J.
  2015{\natexlab{a}}, \mnras, 448, 2987

\bibitem[{{Fosalba} {et~al.}(2015{\natexlab{b}}){Fosalba}, {Gazta{\~n}aga},
  {Castander}, \& {Crocce}}]{Fosalba2015b}
{Fosalba}, P., {Gazta{\~n}aga}, E., {Castander}, F.~J., \& {Crocce}, M.
  2015{\natexlab{b}}, \mnras, 447, 1319

\bibitem[{{Fosalba} {et~al.}(2008){Fosalba}, {Gazta{\~n}aga}, {Castander}, \&
  {Manera}}]{Fosalba2008}
{Fosalba}, P., {Gazta{\~n}aga}, E., {Castander}, F.~J., \& {Manera}, M. 2008,
  \mnras, 391, 435

\bibitem[{{Fry} \& {Gaztanaga}(1993)}]{Fry1993}
{Fry}, J.~N. \& {Gaztanaga}, E. 1993, \apj, 413, 447

\bibitem[{{Gil-Mar{\'\i}n} {et~al.}(2017){Gil-Mar{\'\i}n}, {Percival}, {Verde},
  {Brownstein}, {Chuang}, {Kitaura}, {Rodr{\'\i}guez-Torres}, \&
  {Olmstead}}]{GilMarin2017}
{Gil-Mar{\'\i}n}, H., {Percival}, W.~J., {Verde}, L., {et~al.} 2017, \mnras,
  465, 1757

\bibitem[{{Griffith} {et~al.}(2012){Griffith}, {Cooper}, {Newman}, {Moustakas},
  {Stern}, {Comerford}, {Davis}, {Lotz}, {Barden}, {Conselice}, {Capak},
  {Faber}, {Kirkpatrick}, {Koekemoer}, {Koo}, {Noeske}, {Scoville}, {Sheth},
  {Shopbell}, {Willmer}, \& {Weiner}}]{Griffith2012}
{Griffith}, R.~L., {Cooper}, M.~C., {Newman}, J.~A., {et~al.} 2012, \apjs, 200,
  9

\bibitem[{{Hartlap} {et~al.}(2007){Hartlap}, {Simon}, \&
  {Schneider}}]{Hartlap2007}
{Hartlap}, J., {Simon}, P., \& {Schneider}, P. 2007, \aap, 464, 399

\bibitem[{{Hirata} {et~al.}(2007){Hirata}, {Mandelbaum}, {Ishak}, {Seljak},
  {Nichol}, {Pimbblet}, {Ross}, \& {Wake}}]{Hirata2007}
{Hirata}, C.~M., {Mandelbaum}, R., {Ishak}, M., {et~al.} 2007, \mnras, 381,
  1197

\bibitem[{{Hirata} \& {Seljak}(2004)}]{Hirata2004}
{Hirata}, C.~M. \& {Seljak}, U. 2004, \prd, 70, 063526

\bibitem[{{Hoffmann} {et~al.}(2015){Hoffmann}, {Bel}, {Gazta{\~n}aga},
  {Crocce}, {Fosalba}, \& {Castander}}]{Hoffmann2015}
{Hoffmann}, K., {Bel}, J., {Gazta{\~n}aga}, E., {et~al.} 2015, \mnras, 447,
  1724

\bibitem[{{Hoffmann} {et~al.}(2022){Hoffmann}, {Secco}, {Blazek}, {Crocce},
  {Tallada-Cresp{\'\i}}, {Samuroff}, {Prat}, {Carretero}, {Fosalba},
  {Gazta{\~n}aga}, {Castander}, \& {DES Collaboration}}]{Hoffmann2022}
{Hoffmann}, K., {Secco}, L.~F., {Blazek}, J., {et~al.} 2022, \prd, 106, 123510

\bibitem[{{Ivanov} {et~al.}(2023){Ivanov}, {Philcox}, {Cabass}, {Nishimichi},
  {Simonovi{\'c}}, \& {Zaldarriaga}}]{Ivanov2023}
{Ivanov}, M.~M., {Philcox}, O. H.~E., {Cabass}, G., {et~al.} 2023, \prd, 107,
  083515

\bibitem[{{Ivezi{\'c}} {et~al.}(2019){Ivezi{\'c}}, {Kahn}, {Tyson}, {Abel},
  {Acosta}, {Allsman}, {Alonso}, {AlSayyad}, {Anderson}, {Andrew}, {Angel},
  {Angeli}, {Ansari}, {Antilogus}, {Araujo}, {Armstrong}, {Arndt}, {Astier},
  {Aubourg}, {Auza}, {Axelrod}, {Bard}, {Barr}, {Barrau}, {Bartlett}, {Bauer},
  {Bauman}, {Baumont}, {Bechtol}, {Bechtol}, {Becker}, {Becla}, {Beldica},
  {Bellavia}, {Bianco}, {Biswas}, {Blanc}, {Blazek}, {Blandford}, {Bloom},
  {Bogart}, {Bond}, {Booth}, {Borgland}, {Borne}, {Bosch}, {Boutigny},
  {Brackett}, {Bradshaw}, {Brandt}, {Brown}, {Bullock}, {Burchat}, {Burke},
  {Cagnoli}, {Calabrese}, {Callahan}, {Callen}, {Carlin}, {Carlson},
  {Chandrasekharan}, {Charles-Emerson}, {Chesley}, {Cheu}, {Chiang}, {Chiang},
  {Chirino}, {Chow}, {Ciardi}, {Claver}, {Cohen-Tanugi}, {Cockrum}, {Coles},
  {Connolly}, {Cook}, {Cooray}, {Covey}, {Cribbs}, {Cui}, {Cutri}, {Daly},
  {Daniel}, {Daruich}, {Daubard}, {Daues}, {Dawson}, {Delgado}, {Dellapenna},
  {de Peyster}, {de Val-Borro}, {Digel}, {Doherty}, {Dubois},
  {Dubois-Felsmann}, {Durech}, {Economou}, {Eifler}, {Eracleous}, {Emmons},
  {Fausti Neto}, {Ferguson}, {Figueroa}, {Fisher-Levine}, {Focke}, {Foss},
  {Frank}, {Freemon}, {Gangler}, {Gawiser}, {Geary}, {Gee}, {Geha}, {Gessner},
  {Gibson}, {Gilmore}, {Glanzman}, {Glick}, {Goldina}, {Goldstein}, {Goodenow},
  {Graham}, {Gressler}, {Gris}, {Guy}, {Guyonnet}, {Haller}, {Harris},
  {Hascall}, {Haupt}, {Hernandez}, {Herrmann}, {Hileman}, {Hoblitt}, {Hodgson},
  {Hogan}, {Howard}, {Huang}, {Huffer}, {Ingraham}, {Innes}, {Jacoby}, {Jain},
  {Jammes}, {Jee}, {Jenness}, {Jernigan}, {Jevremovi{\'c}}, {Johns}, {Johnson},
  {Johnson}, {Jones}, {Juramy-Gilles}, {Juri{\'c}}, {Kalirai}, {Kallivayalil},
  {Kalmbach}, {Kantor}, {Karst}, {Kasliwal}, {Kelly}, {Kessler}, {Kinnison},
  {Kirkby}, {Knox}, {Kotov}, {Krabbendam}, {Krughoff}, {Kub{\'a}nek},
  {Kuczewski}, {Kulkarni}, {Ku}, {Kurita}, {Lage}, {Lambert}, {Lange},
  {Langton}, {Le Guillou}, {Levine}, {Liang}, {Lim}, {Lintott}, {Long},
  {Lopez}, {Lotz}, {Lupton}, {Lust}, {MacArthur}, {Mahabal}, {Mandelbaum},
  {Markiewicz}, {Marsh}, {Marshall}, {Marshall}, {May}, {McKercher}, {McQueen},
  {Meyers}, {Migliore}, {Miller}, {Mills}, {Miraval}, {Moeyens}, {Moolekamp},
  {Monet}, {Moniez}, {Monkewitz}, {Montgomery}, {Morrison}, {Mueller},
  {Muller}, {Mu{\~n}oz Arancibia}, {Neill}, {Newbry}, {Nief}, {Nomerotski},
  {Nordby}, {O'Connor}, {Oliver}, {Olivier}, {Olsen}, {O'Mullane}, {Ortiz},
  {Osier}, {Owen}, {Pain}, {Palecek}, {Parejko}, {Parsons}, {Pease},
  {Peterson}, {Peterson}, {Petravick}, {Libby Petrick}, {Petry},
  {Pierfederici}, {Pietrowicz}, {Pike}, {Pinto}, {Plante}, {Plate}, {Plutchak},
  {Price}, {Prouza}, {Radeka}, {Rajagopal}, {Rasmussen}, {Regnault}, {Reil},
  {Reiss}, {Reuter}, {Ridgway}, {Riot}, {Ritz}, {Robinson}, {Roby}, {Roodman},
  {Rosing}, {Roucelle}, {Rumore}, {Russo}, {Saha}, {Sassolas}, {Schalk},
  {Schellart}, {Schindler}, {Schmidt}, {Schneider}, {Schneider}, {Schoening},
  {Schumacher}, {Schwamb}, {Sebag}, {Selvy}, {Sembroski}, {Seppala}, {Serio},
  {Serrano}, {Shaw}, {Shipsey}, {Sick}, {Silvestri}, {Slater}, {Smith},
  {Smith}, {Sobhani}, {Soldahl}, {Storrie-Lombardi}, {Stover}, {Strauss},
  {Street}, {Stubbs}, {Sullivan}, {Sweeney}, {Swinbank}, {Szalay}, {Takacs},
  {Tether}, {Thaler}, {Thayer}, {Thomas}, {Thornton}, {Thukral}, {Tice},
  {Trilling}, {Turri}, {Van Berg}, {Vanden Berk}, {Vetter}, {Virieux},
  {Vucina}, {Wahl}, {Walkowicz}, {Walsh}, {Walter}, {Wang}, {Wang}, {Warner},
  {Wiecha}, {Willman}, {Winters}, {Wittman}, {Wolff}, {Wood-Vasey}, {Wu},
  {Xin}, {Yoachim}, \& {Zhan}}]{Ivezic2019}
{Ivezi{\'c}}, {\v{Z}}., {Kahn}, S.~M., {Tyson}, J.~A., {et~al.} 2019, \apj,
  873, 111

\bibitem[{{Jarvis} {et~al.}(2004){Jarvis}, {Bernstein}, \& {Jain}}]{Jarvis2004}
{Jarvis}, M., {Bernstein}, G., \& {Jain}, B. 2004, \mnras, 352, 338

\bibitem[{{Joachimi} {et~al.}(2015){Joachimi}, {Cacciato}, {Kitching},
  {Leonard}, {Mandelbaum}, {Sch{\"a}fer}, {Sif{\'o}n}, {Hoekstra}, {Kiessling},
  {Kirk}, \& {Rassat}}]{Joachimi2015}
{Joachimi}, B., {Cacciato}, M., {Kitching}, T.~D., {et~al.} 2015, \ssr, 193, 1

\bibitem[{{Joachimi} {et~al.}(2011){Joachimi}, {Mandelbaum}, {Abdalla}, \&
  {Bridle}}]{Joachimi2011}
{Joachimi}, B., {Mandelbaum}, R., {Abdalla}, F.~B., \& {Bridle}, S.~L. 2011,
  \aap, 527, A26

\bibitem[{{Johnston} {et~al.}(2019){Johnston}, {Georgiou}, {Joachimi},
  {Hoekstra}, {Chisari}, {Farrow}, {Fortuna}, {Heymans}, {Joudaki}, {Kuijken},
  \& {Wright}}]{Johnston2019}
{Johnston}, H., {Georgiou}, C., {Joachimi}, B., {et~al.} 2019, \aap, 624, A30

\bibitem[{{Johnston} {et~al.}(2021){Johnston}, {Joachimi}, {Norberg},
  {Hoekstra}, {Eriksen}, {Fortuna}, {Manzoni}, {Serrano}, {Siudek},
  {Tortorelli}, {Asorey}, {Cabayol}, {Carretero}, {Casas}, {Castander},
  {Crocce}, {Fernandez}, {Garc{\'\i}a-Bellido}, {Gaztanaga}, {Hildebrandt},
  {Miquel}, {Navarro-Girones}, {Padilla}, {Sanchez}, {Sevilla-Noarbe}, \&
  {Tallada-Cresp{\'\i}}}]{Johnston2021}
{Johnston}, H., {Joachimi}, B., {Norberg}, P., {et~al.} 2021, \aap, 646, A147

\bibitem[{{Kaiser} \& {Squires}(1993)}]{Kaiser1993}
{Kaiser}, N. \& {Squires}, G. 1993, \apj, 404, 441

\bibitem[{{Kiessling} {et~al.}(2015){Kiessling}, {Cacciato}, {Joachimi},
  {Kirk}, {Kitching}, {Leonard}, {Mandelbaum}, {Sch{\"a}fer}, {Sif{\'o}n},
  {Brown}, \& {Rassat}}]{Kiessling2015}
{Kiessling}, A., {Cacciato}, M., {Joachimi}, B., {et~al.} 2015, \ssr, 193, 67

\bibitem[{{Kirk} {et~al.}(2015){Kirk}, {Brown}, {Hoekstra}, {Joachimi},
  {Kitching}, {Mandelbaum}, {Sif{\'o}n}, {Cacciato}, {Choi}, {Kiessling},
  {Leonard}, {Rassat}, \& {Sch{\"a}fer}}]{Kirk2015}
{Kirk}, D., {Brown}, M.~L., {Hoekstra}, H., {et~al.} 2015, \ssr, 193, 139

\bibitem[{{Kuijken} {et~al.}(2015){Kuijken}, {Heymans}, {Hildebrandt},
  {Nakajima}, {Erben}, {de Jong}, {Viola}, {Choi}, {Hoekstra}, {Miller}, {van
  Uitert}, {Amon}, {Blake}, {Brouwer}, {Buddendiek}, {Conti}, {Eriksen},
  {Grado}, {Harnois-D{\'e}raps}, {Helmich}, {Herbonnet}, {Irisarri},
  {Kitching}, {Klaes}, {La Barbera}, {Napolitano}, {Radovich}, {Schneider},
  {Sif{\'o}n}, {Sikkema}, {Simon}, {Tudorica}, {Valentijn}, {Verdoes Kleijn},
  \& {van Waerbeke}}]{Kuijken2015}
{Kuijken}, K., {Heymans}, C., {Hildebrandt}, H., {et~al.} 2015, \mnras, 454,
  3500

\bibitem[{{Laigle} {et~al.}(2016){Laigle}, {McCracken}, {Ilbert}, {Hsieh},
  {Davidzon}, {Capak}, {Hasinger}, {Silverman}, {Pichon}, {Coupon}, {Aussel},
  {Le Borgne}, {Caputi}, {Cassata}, {Chang}, {Civano}, {Dunlop}, {Fynbo},
  {Kartaltepe}, {Koekemoer}, {Le F{\`e}vre}, {Le Floc'h}, {Leauthaud}, {Lilly},
  {Lin}, {Marchesi}, {Milvang-Jensen}, {Salvato}, {Sanders}, {Scoville},
  {Smolcic}, {Stockmann}, {Taniguchi}, {Tasca}, {Toft}, {Vaccari}, \&
  {Zabl}}]{Laigle2016}
{Laigle}, C., {McCracken}, H.~J., {Ilbert}, O., {et~al.} 2016, \apjs, 224, 24

\bibitem[{{Lamman} {et~al.}(2024){Lamman}, {Tsaprazi}, {Shi},
  {{\v{S}}ar{\v{c}}evi{\'c}}, {Pyne}, {Legnani}, \& {Ferreira}}]{Lamman2024}
{Lamman}, C., {Tsaprazi}, E., {Shi}, J., {et~al.} 2024, The Open Journal of
  Astrophysics, 7, 14

\bibitem[{{Laureijs} {et~al.}(2011){Laureijs}, {Amiaux}, {Arduini},
  {Augu{\`e}res}, {Brinchmann}, {Cole}, {Cropper}, {Dabin}, {Duvet}, {Ealet},
  {Garilli}, {Gondoin}, {Guzzo}, {Hoar}, {Hoekstra}, {Holmes}, {Kitching},
  {Maciaszek}, {Mellier}, {Pasian}, {Percival}, {Rhodes}, {Saavedra Criado},
  {Sauvage}, {Scaramella}, {Valenziano}, {Warren}, {Bender}, {Castander},
  {Cimatti}, {Le F{\`e}vre}, {Kurki-Suonio}, {Levi}, {Lilje}, {Meylan},
  {Nichol}, {Pedersen}, {Popa}, {Rebolo Lopez}, {Rix}, {Rottgering},
  {Zeilinger}, {Grupp}, {Hudelot}, {Massey}, {Meneghetti}, {Miller}, {Paltani},
  {Paulin-Henriksson}, {Pires}, {Saxton}, {Schrabback}, {Seidel}, {Walsh},
  {Aghanim}, {Amendola}, {Bartlett}, {Baccigalupi}, {Beaulieu}, {Benabed},
  {Cuby}, {Elbaz}, {Fosalba}, {Gavazzi}, {Helmi}, {Hook}, {Irwin}, {Kneib},
  {Kunz}, {Mannucci}, {Moscardini}, {Tao}, {Teyssier}, {Weller}, {Zamorani},
  {Zapatero Osorio}, {Boulade}, {Foumond}, {Di Giorgio}, {Guttridge}, {James},
  {Kemp}, {Martignac}, {Spencer}, {Walton}, {Bl{\"u}mchen}, {Bonoli},
  {Bortoletto}, {Cerna}, {Corcione}, {Fabron}, {Jahnke}, {Ligori}, {Madrid},
  {Martin}, {Morgante}, {Pamplona}, {Prieto}, {Riva}, {Toledo}, {Trifoglio},
  {Zerbi}, {Abdalla}, {Douspis}, {Grenet}, {Borgani}, {Bouwens}, {Courbin},
  {Delouis}, {Dubath}, {Fontana}, {Frailis}, {Grazian}, {Koppenh{\"o}fer},
  {Mansutti}, {Melchior}, {Mignoli}, {Mohr}, {Neissner}, {Noddle}, {Poncet},
  {Scodeggio}, {Serrano}, {Shane}, {Starck}, {Surace}, {Taylor},
  {Verdoes-Kleijn}, {Vuerli}, {Williams}, {Zacchei}, {Altieri}, {Escudero
  Sanz}, {Kohley}, {Oosterbroek}, {Astier}, {Bacon}, {Bardelli}, {Baugh},
  {Bellagamba}, {Benoist}, {Bianchi}, {Biviano}, {Branchini}, {Carbone},
  {Cardone}, {Clements}, {Colombi}, {Conselice}, {Cresci}, {Deacon}, {Dunlop},
  {Fedeli}, {Fontanot}, {Franzetti}, {Giocoli}, {Garcia-Bellido}, {Gow},
  {Heavens}, {Hewett}, {Heymans}, {Holland}, {Huang}, {Ilbert}, {Joachimi},
  {Jennins}, {Kerins}, {Kiessling}, {Kirk}, {Kotak}, {Krause}, {Lahav}, {van
  Leeuwen}, {Lesgourgues}, {Lombardi}, {Magliocchetti}, {Maguire}, {Majerotto},
  {Maoli}, {Marulli}, {Maurogordato}, {McCracken}, {McLure}, {Melchiorri},
  {Merson}, {Moresco}, {Nonino}, {Norberg}, {Peacock}, {Pello}, {Penny},
  {Pettorino}, {Di Porto}, {Pozzetti}, {Quercellini}, {Radovich}, {Rassat},
  {Roche}, {Ronayette}, {Rossetti}, {Sartoris}, {Schneider}, {Semboloni},
  {Serjeant}, {Simpson}, {Skordis}, {Smadja}, {Smartt}, {Spano}, {Spiro},
  {Sullivan}, {Tilquin}, {Trotta}, {Verde}, {Wang}, {Williger}, {Zhao},
  {Zoubian}, \& {Zucca}}]{Laureijs2011}
{Laureijs}, R., {Amiaux}, J., {Arduini}, S., {et~al.} 2011, arXiv e-prints,
  arXiv:1110.3193

\bibitem[{{Li} {et~al.}(2023){Li}, {Zhang}, {Sugiyama}, {Dalal}, {Terasawa},
  {Rau}, {Mandelbaum}, {Takada}, {More}, {Strauss}, {Miyatake}, {Shirasaki},
  {Hamana}, {Oguri}, {Luo}, {Nishizawa}, {Takahashi}, {Nicola}, {Osato},
  {Kannawadi}, {Sunayama}, {Armstrong}, {Bosch}, {Komiyama}, {Lupton}, {Lust},
  {MacArthur}, {Miyazaki}, {Murayama}, {Nishimichi}, {Okura}, {Price}, {Tait},
  {Tanaka}, \& {Wang}}]{Li2023}
{Li}, X., {Zhang}, T., {Sugiyama}, S., {et~al.} 2023, \prd, 108, 123518

\bibitem[{{Linke} {et~al.}(2020{\natexlab{a}}){Linke}, {Simon}, {Schneider},
  {Erben}, {Farrow}, {Heymans}, {Hildebrandt}, {Hopkins}, {Kannawadi},
  {Napolitano}, {Sif{\'o}n}, \& {Wright}}]{Linke2020b}
{Linke}, L., {Simon}, P., {Schneider}, P., {et~al.} 2020{\natexlab{a}}, \aap,
  640, A59

\bibitem[{{Linke} {et~al.}(2022){Linke}, {Simon}, {Schneider}, {Farrow},
  {R{\"o}diger}, \& {Wright}}]{Linke2022}
{Linke}, L., {Simon}, P., {Schneider}, P., {et~al.} 2022, \aap, 665, A38

\bibitem[{{Linke} {et~al.}(2020{\natexlab{b}}){Linke}, {Simon}, {Schneider}, \&
  {Hilbert}}]{Linke2020a}
{Linke}, L., {Simon}, P., {Schneider}, P., \& {Hilbert}, S. 2020{\natexlab{b}},
  \aap, 634, A13

\bibitem[{{Maion} {et~al.}(2024){Maion}, {Angulo}, {Bakx}, {Chisari}, {Kurita},
  \& {Pellejero-Ib{\'a}{\~n}ez}}]{Maion2024}
{Maion}, F., {Angulo}, R.~E., {Bakx}, T., {et~al.} 2024, \mnras, 531, 2684

\bibitem[{{Mar{\'\i}n} {et~al.}(2013){Mar{\'\i}n}, {Blake}, {Poole}, {McBride},
  {Brough}, {Colless}, {Contreras}, {Couch}, {Croton}, {Croom}, {Davis},
  {Drinkwater}, {Forster}, {Gilbank}, {Gladders}, {Glazebrook}, {Jelliffe},
  {Jurek}, {Li}, {Madore}, {Martin}, {Pimbblet}, {Pracy}, {Sharp}, {Wisnioski},
  {Woods}, {Wyder}, \& {Yee}}]{Marin2013}
{Mar{\'\i}n}, F.~A., {Blake}, C., {Poole}, G.~B., {et~al.} 2013, \mnras, 432,
  2654

\bibitem[{Nelder \& Mead(1965)}]{Nelder1965}
Nelder, J.~A. \& Mead, R. 1965, The Computer Journal, 7, 308

\bibitem[{{Planck Collaboration: Aghanim} {et~al.}(2020){Planck Collaboration:
  Aghanim}, {Akrami}, {Ashdown}, {Aumont}, {Baccigalupi}, {Ballardini},
  {Banday}, {Barreiro}, {Bartolo}, {Basak}, {Battye}, {Benabed}, {Bernard},
  {Bersanelli}, {Bielewicz}, {Bock}, {Bond}, {Borrill}, {Bouchet}, {Boulanger},
  {Bucher}, {Burigana}, {Butler}, {Calabrese}, {Cardoso}, {Carron},
  {Challinor}, {Chiang}, {Chluba}, {Colombo}, {Combet}, {Contreras}, {Crill},
  {Cuttaia}, {de Bernardis}, {de Zotti}, {Delabrouille}, {Delouis}, {Di
  Valentino}, {Diego}, {Dor{\'e}}, {Douspis}, {Ducout}, {Dupac}, {Dusini},
  {Efstathiou}, {Elsner}, {En{\ss}lin}, {Eriksen}, {Fantaye}, {Farhang},
  {Fergusson}, {Fernandez-Cobos}, {Finelli}, {Forastieri}, {Frailis},
  {Fraisse}, {Franceschi}, {Frolov}, {Galeotta}, {Galli}, {Ganga},
  {G{\'e}nova-Santos}, {Gerbino}, {Ghosh}, {Gonz{\'a}lez-Nuevo}, {G{\'o}rski},
  {Gratton}, {Gruppuso}, {Gudmundsson}, {Hamann}, {Handley}, {Hansen},
  {Herranz}, {Hildebrandt}, {Hivon}, {Huang}, {Jaffe}, {Jones}, {Karakci},
  {Keih{\"a}nen}, {Keskitalo}, {Kiiveri}, {Kim}, {Kisner}, {Knox},
  {Krachmalnicoff}, {Kunz}, {Kurki-Suonio}, {Lagache}, {Lamarre}, {Lasenby},
  {Lattanzi}, {Lawrence}, {Le Jeune}, {Lemos}, {Lesgourgues}, {Levrier},
  {Lewis}, {Liguori}, {Lilje}, {Lilley}, {Lindholm}, {L{\'o}pez-Caniego},
  {Lubin}, {Ma}, {Mac{\'\i}as-P{\'e}rez}, {Maggio}, {Maino}, {Mandolesi},
  {Mangilli}, {Marcos-Caballero}, {Maris}, {Martin}, {Martinelli},
  {Mart{\'\i}nez-Gonz{\'a}lez}, {Matarrese}, {Mauri}, {McEwen}, {Meinhold},
  {Melchiorri}, {Mennella}, {Migliaccio}, {Millea}, {Mitra},
  {Miville-Desch{\^e}nes}, {Molinari}, {Montier}, {Morgante}, {Moss}, {Natoli},
  {N{\o}rgaard-Nielsen}, {Pagano}, {Paoletti}, {Partridge}, {Patanchon},
  {Peiris}, {Perrotta}, {Pettorino}, {Piacentini}, {Polastri}, {Polenta},
  {Puget}, {Rachen}, {Reinecke}, {Remazeilles}, {Renzi}, {Rocha}, {Rosset},
  {Roudier}, {Rubi{\~n}o-Mart{\'\i}n}, {Ruiz-Granados}, {Salvati}, {Sandri},
  {Savelainen}, {Scott}, {Shellard}, {Sirignano}, {Sirri}, {Spencer},
  {Sunyaev}, {Suur-Uski}, {Tauber}, {Tavagnacco}, {Tenti}, {Toffolatti},
  {Tomasi}, {Trombetti}, {Valenziano}, {Valiviita}, {Van Tent}, {Vibert},
  {Vielva}, {Villa}, {Vittorio}, {Wandelt}, {Wehus}, {White}, {White},
  {Zacchei}, \& {Zonca}}]{Planck2018}
{Planck Collaboration: Aghanim}, N., {Akrami}, Y., {Ashdown}, M., {et~al.}
  2020, \aap, 641, A6

\bibitem[{{Pyne} \& {Joachimi}(2021)}]{Pyne2021}
{Pyne}, S. \& {Joachimi}, B. 2021, \mnras, 503, 2300

\bibitem[{{Pyne} {et~al.}(2022){Pyne}, {Tenneti}, \& {Joachimi}}]{Pyne2022}
{Pyne}, S., {Tenneti}, A., \& {Joachimi}, B. 2022, \mnras, 516, 1829

\bibitem[{{Reyes} {et~al.}(2012){Reyes}, {Mandelbaum}, {Gunn}, {Nakajima},
  {Seljak}, \& {Hirata}}]{Reyes2012}
{Reyes}, R., {Mandelbaum}, R., {Gunn}, J.~E., {et~al.} 2012, \mnras, 425, 2610

\bibitem[{{Samuroff} {et~al.}(2021){Samuroff}, {Mandelbaum}, \&
  {Blazek}}]{Samuroff2021}
{Samuroff}, S., {Mandelbaum}, R., \& {Blazek}, J. 2021, \mnras, 508, 637

\bibitem[{{Samuroff} {et~al.}(2023){Samuroff}, {Mandelbaum}, {Blazek},
  {Campos}, {MacCrann}, {Zacharegkas}, {Amon}, {Prat}, {Singh}, {Elvin-Poole},
  {Ross}, {Alarcon}, {Baxter}, {Bechtol}, {Becker}, {Bernstein}, {Rosell},
  {Kind}, {Cawthon}, {Chang}, {Chen}, {Choi}, {Crocce}, {Davis}, {DeRose},
  {Dodelson}, {Doux}, {Drlica-Wagner}, {Eckert}, {Everett}, {Fert{\'e}},
  {Gatti}, {Giannini}, {Gruen}, {Gruendl}, {Harrison}, {Herner}, {Huff},
  {Jarvis}, {Kuropatkin}, {Leget}, {Lemos}, {McCullough}, {Myles},
  {Navarro-Alsina}, {Pandey}, {Porredon}, {Raveri}, {Rodriguez-Monroy},
  {Rollins}, {Roodman}, {Rossi}, {Rykoff}, {S{\'a}nchez}, {Secco},
  {Sevilla-Noarbe}, {Sheldon}, {Shin}, {Troxel}, {Tutusaus}, {Weaverdyck},
  {Yanny}, {Yin}, {Zhang}, {Zuntz}, {Aguena}, {Alves}, {Annis}, {Bacon},
  {Bertin}, {Bocquet}, {Brooks}, {Burke}, {Carretero}, {Costanzi}, {da Costa},
  {Pereira}, {De Vicente}, {Desai}, {Diehl}, {Dietrich}, {Doel}, {Ferrero},
  {Flaugher}, {Frieman}, {Garc{\'\i}a-Bellido}, {Hinton}, {Hollowood},
  {Honscheid}, {James}, {Kuehn}, {Lahav}, {Marshall}, {Melchior},
  {Mena-Fern{\'a}ndez}, {Menanteau}, {Miquel}, {Newman}, {Palmese}, {Pieres},
  {Malag{\'o}n}, {Sanchez}, {Scarpine}, {Smith}, {Suchyta}, {Swanson}, {Tarle},
  {To}, \& {DES Collaboration}}]{Samuroff2023}
{Samuroff}, S., {Mandelbaum}, R., {Blazek}, J., {et~al.} 2023, \mnras, 524,
  2195

\bibitem[{{Schneider} \& {Bridle}(2010)}]{Schneider2010}
{Schneider}, M.~D. \& {Bridle}, S. 2010, \mnras, 402, 2127

\bibitem[{{Schneider} \& {Lombardi}(2003)}]{Schneider2003}
{Schneider}, P. \& {Lombardi}, M. 2003, \aap, 397, 809

\bibitem[{{Schneider} {et~al.}(2002){Schneider}, {van Waerbeke}, {Kilbinger},
  \& {Mellier}}]{Schneider2002}
{Schneider}, P., {van Waerbeke}, L., {Kilbinger}, M., \& {Mellier}, Y. 2002,
  \aap, 396, 1

\bibitem[{{Schneider} \& {Watts}(2005)}]{Schneider2005}
{Schneider}, P. \& {Watts}, P. 2005, \aap, 432, 783

\bibitem[{{Scoccimarro} \& {Couchman}(2001)}]{Scoccimarro2001}
{Scoccimarro}, R. \& {Couchman}, H.~M.~P. 2001, \mnras, 325, 1312

\bibitem[{{Secco} {et~al.}(2022){Secco}, {Samuroff}, {Krause}, {Jain},
  {Blazek}, {Raveri}, {Campos}, {Amon}, {Chen}, {Doux}, {Choi}, {Gruen},
  {Bernstein}, {Chang}, {DeRose}, {Myles}, {Fert{\'e}}, {Lemos}, {Huterer},
  {Prat}, {Troxel}, {MacCrann}, {Liddle}, {Kacprzak}, {Fang}, {S{\'a}nchez},
  {Pandey}, {Dodelson}, {Chintalapati}, {Hoffmann}, {Alarcon}, {Alves},
  {Andrade-Oliveira}, {Baxter}, {Bechtol}, {Becker}, {Brandao-Souza},
  {Camacho}, {Carnero Rosell}, {Carrasco Kind}, {Cawthon}, {Cordero}, {Crocce},
  {Davis}, {Di Valentino}, {Drlica-Wagner}, {Eckert}, {Eifler}, {Elidaiana},
  {Elsner}, {Elvin-Poole}, {Everett}, {Fosalba}, {Friedrich}, {Gatti},
  {Giannini}, {Gruendl}, {Harrison}, {Hartley}, {Herner}, {Huang}, {Huff},
  {Jarvis}, {Jeffrey}, {Kuropatkin}, {Leget}, {Muir}, {Mccullough}, {Navarro
  Alsina}, {Omori}, {Park}, {Porredon}, {Rollins}, {Roodman}, {Rosenfeld},
  {Ross}, {Rykoff}, {Sanchez}, {Sevilla-Noarbe}, {Sheldon}, {Shin}, {Troja},
  {Tutusaus}, {Varga}, {Weaverdyck}, {Wechsler}, {Yanny}, {Yin}, {Zhang},
  {Zuntz}, {Abbott}, {Aguena}, {Allam}, {Annis}, {Bacon}, {Bertin}, {Bhargava},
  {Bridle}, {Brooks}, {Buckley-Geer}, {Burke}, {Carretero}, {Costanzi}, {da
  Costa}, {De Vicente}, {Diehl}, {Dietrich}, {Doel}, {Ferrero}, {Flaugher},
  {Frieman}, {Garc{\'\i}a-Bellido}, {Gaztanaga}, {Gerdes}, {Giannantonio},
  {Gschwend}, {Gutierrez}, {Hinton}, {Hollowood}, {Honscheid}, {Hoyle},
  {James}, {Jeltema}, {Kuehn}, {Lahav}, {Lima}, {Lin}, {Maia}, {Marshall},
  {Martini}, {Melchior}, {Menanteau}, {Miquel}, {Mohr}, {Morgan}, {Ogando},
  {Palmese}, {Paz-Chinch{\'o}n}, {Petravick}, {Pieres}, {Plazas Malag{\'o}n},
  {Rodriguez-Monroy}, {Romer}, {Sanchez}, {Scarpine}, {Schubnell}, {Scolnic},
  {Serrano}, {Smith}, {Soares-Santos}, {Suchyta}, {Swanson}, {Tarle}, {Thomas},
  {To}, \& {DES Collaboration}}]{Secco2022}
{Secco}, L.~F., {Samuroff}, S., {Krause}, E., {et~al.} 2022, \prd, 105, 023515

\bibitem[{{Semboloni} {et~al.}(2008){Semboloni}, {Heymans}, {van Waerbeke}, \&
  {Schneider}}]{Semboloni2008}
{Semboloni}, E., {Heymans}, C., {van Waerbeke}, L., \& {Schneider}, P. 2008,
  \mnras, 388, 991

\bibitem[{{Simon} {et~al.}(2012){Simon}, {Schneider}, \&
  {K{\"u}bler}}]{Simon2012}
{Simon}, P., {Schneider}, P., \& {K{\"u}bler}, D. 2012, \aap, 548, A102

\bibitem[{{Singh} \& {Mandelbaum}(2016)}]{Singh2016}
{Singh}, S. \& {Mandelbaum}, R. 2016, \mnras, 457, 2301

\bibitem[{{Singh} {et~al.}(2015){Singh}, {Mandelbaum}, \& {More}}]{Singh2015}
{Singh}, S., {Mandelbaum}, R., \& {More}, S. 2015, \mnras, 450, 2195

\bibitem[{{Slepian} {et~al.}(2017){Slepian}, {Eisenstein}, {Brownstein},
  {Chuang}, {Gil-Mar{\'\i}n}, {Ho}, {Kitaura}, {Percival}, {Ross}, {Rossi},
  {Seo}, {Slosar}, \& {Vargas-Maga{\~n}a}}]{Slepian2017}
{Slepian}, Z., {Eisenstein}, D.~J., {Brownstein}, J.~R., {et~al.} 2017, \mnras,
  469, 1738

\bibitem[{{Takada} \& {Jain}(2003)}]{Takada2003}
{Takada}, M. \& {Jain}, B. 2003, \mnras, 340, 580

\bibitem[{{Takahashi} {et~al.}(2020){Takahashi}, {Nishimichi}, {Namikawa},
  {Taruya}, {Kayo}, {Osato}, {Kobayashi}, \& {Shirasaki}}]{Takahashi2020}
{Takahashi}, R., {Nishimichi}, T., {Namikawa}, T., {et~al.} 2020, \apj, 895,
  113

\bibitem[{{Takahashi} {et~al.}(2012){Takahashi}, {Sato}, {Nishimichi},
  {Taruya}, \& {Oguri}}]{Takahashi2012}
{Takahashi}, R., {Sato}, M., {Nishimichi}, T., {Taruya}, A., \& {Oguri}, M.
  2012, \apj, 761, 152

\bibitem[{{Tallada} {et~al.}(2020){Tallada}, {Carretero}, {Casals},
  {Acosta-Silva}, {Serrano}, {Caubet}, {Castander}, {C{\'e}sar}, {Crocce},
  {Delfino}, {Eriksen}, {Fosalba}, {Gazta{\~n}aga}, {Merino}, {Neissner}, \&
  {Tonello}}]{Tallada2020}
{Tallada}, P., {Carretero}, J., {Casals}, J., {et~al.} 2020, Astronomy and
  Computing, 32, 100391

\bibitem[{{The LSST Dark Energy Science Collaboration: Mandelbaum}
  {et~al.}(2018){The LSST Dark Energy Science Collaboration: Mandelbaum},
  {Eifler}, {Hlo{\v{z}}ek}, {Collett}, {Gawiser}, {Scolnic}, {Alonso}, {Awan},
  {Biswas}, {Blazek}, {Burchat}, {Chisari}, {Dell'Antonio}, {Digel}, {Frieman},
  {Goldstein}, {Hook}, {Ivezi{\'c}}, {Kahn}, {Kamath}, {Kirkby}, {Kitching},
  {Krause}, {Leget}, {Marshall}, {Meyers}, {Miyatake}, {Newman}, {Nichol},
  {Rykoff}, {Sanchez}, {Slosar}, {Sullivan}, \& {Troxel}}]{LSST2018}
{The LSST Dark Energy Science Collaboration: Mandelbaum}, R., {Eifler}, T.,
  {Hlo{\v{z}}ek}, R., {et~al.} 2018, arXiv e-prints, arXiv:1809.01669

\bibitem[{{Troxel} \& {Ishak}(2015)}]{Troxel2015}
{Troxel}, M.~A. \& {Ishak}, M. 2015, \physrep, 558, 1

\bibitem[{{Vlah} {et~al.}(2020){Vlah}, {Chisari}, \& {Schmidt}}]{Vlah2020}
{Vlah}, Z., {Chisari}, N.~E., \& {Schmidt}, F. 2020, \jcap, 01, 025

\bibitem[{{Yankelevich} \& {Porciani}(2019)}]{Yankelevich2019}
{Yankelevich}, V. \& {Porciani}, C. 2019, \mnras, 483, 2078

\bibitem[{{Zehavi} {et~al.}(2011){Zehavi}, {Zheng}, {Weinberg}, {Blanton},
  {Bahcall}, {Berlind}, {Brinkmann}, {Frieman}, {Gunn}, {Lupton}, {Nichol},
  {Percival}, {Schneider}, {Skibba}, {Strauss}, {Tegmark}, \&
  {York}}]{Zehavi2011}
{Zehavi}, I., {Zheng}, Z., {Weinberg}, D.~H., {et~al.} 2011, \apj, 736, 59

\end{thebibliography}

%

\begin{appendix}
  \onecolumn 
\section{Relation of aperture statistics to bispectrum \label{app: bispectrum}}
  
To model $\NNMIA$, we relate it to the galaxy-galaxy-shape bispectrum $B_\mathrm{ggI}$. For this, we first write it as
\begin{align}
    \notag\NNMIA(R_1, R_2, R_3) &= \int \dd{\chi} p(\chi)\int \dd{\chi_1} p(\chi_1)\int \dd{\chi_2} p(\chi_2)\, \int \dd[2]{r_1'}\int \dd[2]{r_2'}\int \dd[2]{r_3'}\, U_R(r_1')\, U_R(r_2')\, U_R(r_3')\\
    &\quad \times \expval{\deltaGal(\rvec-\rvec_1, \chi_1)\,\deltaGal(\rvec-\rvec_2, \chi_2)\, \deltaIA(\rvec-\rvec_3, \chi)}\;.
\end{align}
We now use the galaxy-galaxy-intrinsic shape bispectrum $B_\mathrm{ggI}$ defined by
\begin{align}
    (2\pi)^3 \dirac(\kvec_1+\kvec_2+\kvec_3)\, B_\mathrm{ggI}(k_1, k_2, k_3; \chi_1, \chi_2, \chi)&= \expval{\fourier{\deltaGal}(\kvec_1, \chi_1)\, \fourier{\deltaGal}(\kvec_2, \chi_2)\, \fourier{\deltaIA}(\kvec_3, \chi)}\;,
\end{align}
and separate the $\Vec{k}_i$ into their components $k_{z,i}$ along the line-of-sight and $\kvec_{\perp, i}$ perpendicular to the line-of-sight. With this,
\begin{align}
    \NNMIA(R_1, R_2, R_3) & = \int \dd{\chi} p(\chi)\int \dd{\chi_1} p(\chi_1)\int \dd{\chi_2} p(\chi_2)\,  \int \frac{\dd[2]{k_{\perp, 1}}}{(2\pi)^2}\int \frac{\dd[2]{k_{\perp, 2}}}{(2\pi)^2}\int \frac{\dd[2]{k_{\perp, 3}}}{(2\pi)^2}\, \fourier{U}_R(k_{\perp, 1})\, \fourier{U}_R(k_{\perp, 2})\fourier{U}_R(k_{\perp, 3})\\
    &\notag \quad \times \int \frac{\dd{k_{z,1}}}{2\pi} \int \frac{\dd{k_{z,2}}}{2\pi} \int \frac{\dd{k_{z,3}}}{2\pi} \E^{-\I (k_{z,1}\,\chi_1+k_{z,2}\,\chi_2+k_{z,3}\,\chi}\,  (2\pi)^3 \dirac(\kvec_1+\kvec_2+\kvec_3)\, B_\mathrm{ggI}(k_1, k_2, k_3; \chi_1, \chi_2, \chi)\;.
\end{align}
Evaluating the $k_3$ integrals leads to
\begin{align}
    \NNMIA(R_1, R_2, R_3) &= \int \dd{\chi} p(\chi)\int \dd{\chi_1} p(\chi_1)\int \dd{\chi_2} p(\chi_2)\,  \int \frac{\dd[2]{k_{\perp, 1}}}{(2\pi)^2}\int \frac{\dd[2]{k_{\perp, 2}}}{(2\pi)^2}\, \fourier{U}_R(k_{\perp, 1})\, \fourier{U}_R(k_{\perp, 2})\fourier{U}_R(|k_{\perp, 1}+k_{\perp, 2}|)\\
    &\notag \quad \times \int \frac{\dd{k_{z,1}}}{2\pi} \int \frac{\dd{k_{z,2}}}{2\pi}  \E^{-\I k_{z,1}\,(\chi_1-\chi)+\I k_{z,2}\,(\chi_2-\chi)}\,  B_\mathrm{ggI}(k_1, k_2, k_3; \chi_1, \chi_2, \chi)\;.
\end{align}
Now, we use the Limber approximation, under which
\begin{equation}
B_\mathrm{ggI}(k_1, k_2, k_3; \chi_1, \chi_2, \chi) \simeq B_\mathrm{ggI}(k_{\perp, 1}, k_{\perp, 2}, k_{\perp, 3}; \chi_1, \chi_2, \chi)\;.    
\end{equation}
Inserting this, evaluating the $k_{z,i}$ integrals and the (then trivial) $\chi_1$ and $\chi_2$ integrals leads to
\begin{align}
    &\NNMIA(R_1, R_2, R_3)=   \int \frac{\dd[2]{k_{\perp, 1}}}{(2\pi)^2}\int \frac{\dd[2]{k_{\perp, 2}}}{(2\pi)^2}\, \fourier{U}_R(k_{\perp, 1})\, \fourier{U}_R(k_{\perp, 2})\fourier{U}_R(|k_{\perp, 1}+k_{\perp, 2}|)\, \int \dd{\chi} p^3(\chi) B_\mathrm{ggI}(k_{\perp, 1}, k_{\perp, 2}, k_{\perp, 3}; \chi, \chi, \chi)\;,
\end{align}
so $\NNMIA$ can be readily computed for a given filter function $U$, a galaxy redshift distribution $p$ and a bispectrum model $B_\mathrm{ggI}$.

\section{Impact of non-linear galaxy bias \label{app: nonlinear}}
\begin{figure}
    \centering
    \includegraphics[width=0.5\linewidth]{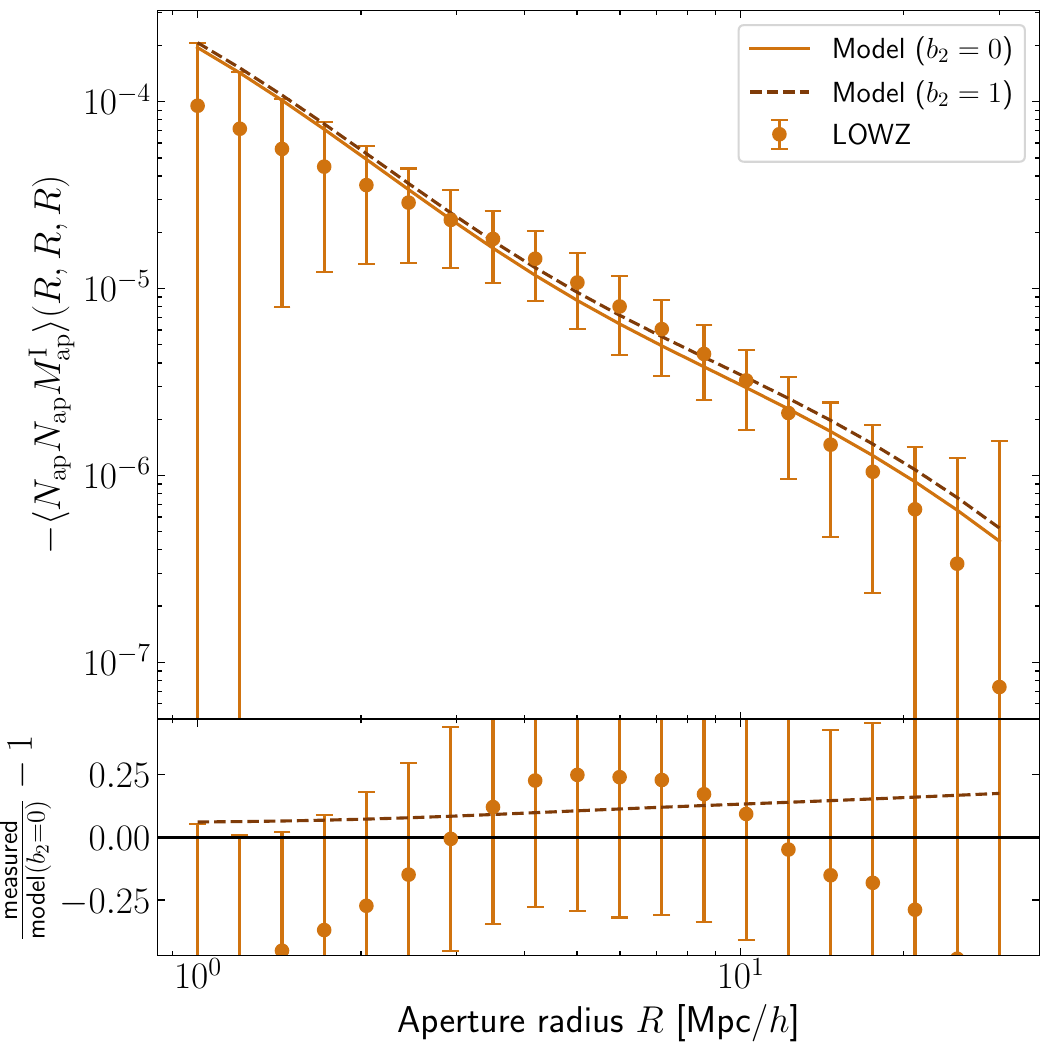}
    \caption{Comparison of IA measurements for LOWZ sample with and without non-linear bias. Upper panel: Measurement of $\NNMIA$ for full LOWZ sample (points), with model fit without non-linear bias (solid line) and same model but using $b_2=1$ (dashed line). Lower panel: Relative difference to the model, which neglects non-linear bias.}
    \label{fig: nonlinear}
\end{figure}

As described in Eq.~\eqref{eq: bispec}, the galaxy-matter bispectrum already depends on the non-linear galaxy bias $b_2$ at the leading order. However, we argue that $b_2$ can be neglected for the $\NNMIA$ modelling at the level of uncertainty of the LOWZ measurements. To demonstrate this, Fig.~\ref{fig: nonlinear} shows the best-fitting model for the LOWZ measurements in the full sample when neglecting $b_2$ and the model for the same $A_\mathrm{IA}$  and $b$ but now setting $b_2=1$. For this, we computed the non-linear power spectrum using the revised \verb|halofit| prescription \citep{Takahashi2012}. 

When including a positive $b_2$, the model increases, particularly at larger scales, by up to $20\%$. However, this increase is small compared to the measurement uncertainty and both models show good agreement with the measurement. Therefore, the value of $b_2$ is not critical to describe the $\NNMIA$ from the LOWZ. 

We note that Fig~\ref{fig: nonlinear} depicts a pessimistic case, as we do not expect $b_2$ to be as large as one. Measurements of the bispectrum of galaxies from the LOWZ sample by \citet{GilMarin2017} find $b_2\,\sigma_8=0.6$, which for $\sigma_8=0.8$ yields $b_2=0.75$. Using LOWZ-like simulated mock galaxies, \citet{Eggemeier2021} found the even lower value of $b_2=0.3\pm0.2$. Consequently, the impact of the non-linear galaxy bias is likely even smaller than shown here.

\section{Estimator bias \label{app: bias}}
The estimator for the correlation function in Eq.~\eqref{eq: biased estimator} is biased. To see this, we calculate the expectation value of the estimator. The expectation value of the numerator is
\begin{align}
    &\expval{\sum_{i=1}^{N_\mathrm{L}}\sum_{j=1}^{N_\mathrm{L}}\sum_{k=1}^{N_\mathrm{S}}  \eps_k\,\E^{\I (\phi_i+\phi_j)} \triangle(r_1, r_2, \phi, \Vec{x}_i, \Vec{x}_j, \Vec{x}_k)\, \Theta_{\Pimax}(\chi_i, \chi_j; \chi_k)}\\
    &= \int \dd[2]{x_1} \int \dd[2]{x_2} \int \dd[2]{x_3} \int \dd{\chi_1} \int \dd{\chi_2} \int \dd{\chi_3} \,  p(\chi_1)\, p(\chi_2)\, p(\chi_3)\, \expval{n(\Vec{x}_1, \chi_1)\, n(\Vec{x}_2, \chi_2)\, \epsChi(\Vec{x}_3, \chi_3)\, \E^{-\I(\phi_1+\phi_2)}} \\
    &\notag\quad \times \triangle(r_1, r_2, \phi, \Vec{x}_1, \Vec{x}_2, \Vec{x}_3)\,  \Theta_{\Pimax}(\chi_1, \chi_2; \chi_3)\\
    &=\int \dd{\chi} \int_{\chi-\Pimax}^{\chi+\Pimax} \dd{\chi_1} \int_{\chi-\Pimax}^{\chi+\Pimax} \dd{\chi_2}  p(\chi_1)\, p(\chi_2)\, p(\chi)\,  \expval{n(\Vec{x}_3+\rvec_1, \chi_1)\, n(\Vec{x}_3+\rvec_2, \chi_2)\, \epsChi(\Vec{x}_3, \chi)\, \E^{-\I(\phi_1+\phi_2)}} \\
    &\notag = \aveNumDenTwoD^2_{\Pimax} \GGGL^{\Pimax}(\rvec_1, \rvec_2) \;.
\end{align}
The expectation value of the denominator is
\begin{align}
    &\expval{\sum_{i=1}^{N_\mathrm{L}}\sum_{j=1}^{N_\mathrm{L}}\sum_{k=1}^{N_\mathrm{S}}  \triangle(r_1, r_2, \phi, \Vec{x}_i, \Vec{x}_j, \Vec{x}_k)\, \Theta_{\Pimax}(\chi_i, \chi_j;\chi_k)}\\
    &= \int \dd[2]{x_1} \int \dd[2]{x_2} \int \dd[2]{x_3} \int \dd{\chi_1} \int \dd{\chi_2} \int \dd{\chi_3} \, p(\chi_1)\, p(\chi_2)\, p(\chi_3)\, \expval{n(\Vec{x}_1, \chi_1)\, n(\Vec{x}_2, \chi_2)} \, \triangle(r_1, r_2, \phi, \Vec{x}_1, \Vec{x}_2, \Vec{x}_3)\,  \Theta_{\Pimax}(\chi_1, \chi_2, \chi_3)\\
    &=\int \dd{\chi} \int_{\chi-\Pimax}^{\chi+\Pimax} \dd{\chi_1} \int_{\chi-\Pimax}^{\chi+\Pimax} \dd{\chi_2}  p(\chi_1)\, p(\chi_2)\, p(\chi)\,  \expval{n(\Vec{x}_3+\rvec_1, \chi_1)\, n(\Vec{x}_3+\rvec_2, \chi_2)} \\
        &=\int \dd{\chi} \int_{\chi-\Pimax}^{\chi+\Pimax} \dd{\chi_1} \int_{\chi-\Pimax}^{\chi+\Pimax} \dd{\chi_2}  p(\chi_1)\, p(\chi_2)\, p(\chi)\,  \times\aveNumDenThreeD(\chi_1) \, \aveNumDenThreeD(\chi_2) [1+\xi(|\rvec_1-\rvec_2|, |\chi_1-\chi_2|)]\;. 
\end{align}
Here, $\xi(\Delta r, \Delta \chi)$ is the three-dimensional galaxy clustering function for galaxy pairs separated by $\Delta r$ in projection and by $\Delta \chi$ along the line-of-sight.
This expectation value can be approximated by
\begin{align}
&        \int \dd{\chi} \int_{\chi-\Pimax}^{\chi+\Pimax} \dd{\chi_1} \int_{\chi-\Pimax}^{\chi+\Pimax} \dd{\chi_2}  p(\chi_1)\, p(\chi_2)\, p(\chi)\, \aveNumDenThreeD(\chi_1) \, \aveNumDenThreeD(\chi_2) \,[1+\xi(|\rvec_1-\rvec_2|, |\chi_1-\chi_2|)]\\
        &\notag \simeq \aveNumDenTwoD^2_{\Pimax}\,         \int \dd{\chi} \int_{\chi-\Pimax}^{\chi+\Pimax} \dd{\chi_1} \int_{\chi-\Pimax}^{\chi+\Pimax} \dd{\chi_2}  p(\chi_1)\, p(\chi_2)\, p(\chi)\,  [1+\xi(|\rvec_1-\rvec_2|, |\chi_1-\chi_2|)] \\
        &\notag =  \aveNumDenTwoD^2_{\Pimax}\,   B(\rvec_1, \rvec_2, \Pimax)\;.
\end{align}
Therefore, the expectation value of the estimator $\hat{\GGGL}^{\Pimax}_\mathrm{bias}$ is
\begin{align}
    \expval{\hat{\GGGL}^{\Pimax}_\mathrm{bias}} \simeq \frac{ \GGGL^{\Pimax}}{ B(\rvec_1, \rvec_2, \Pimax)}\;.
\end{align}

\begin{figure}
    \centering
    \includegraphics[width=0.5\linewidth]{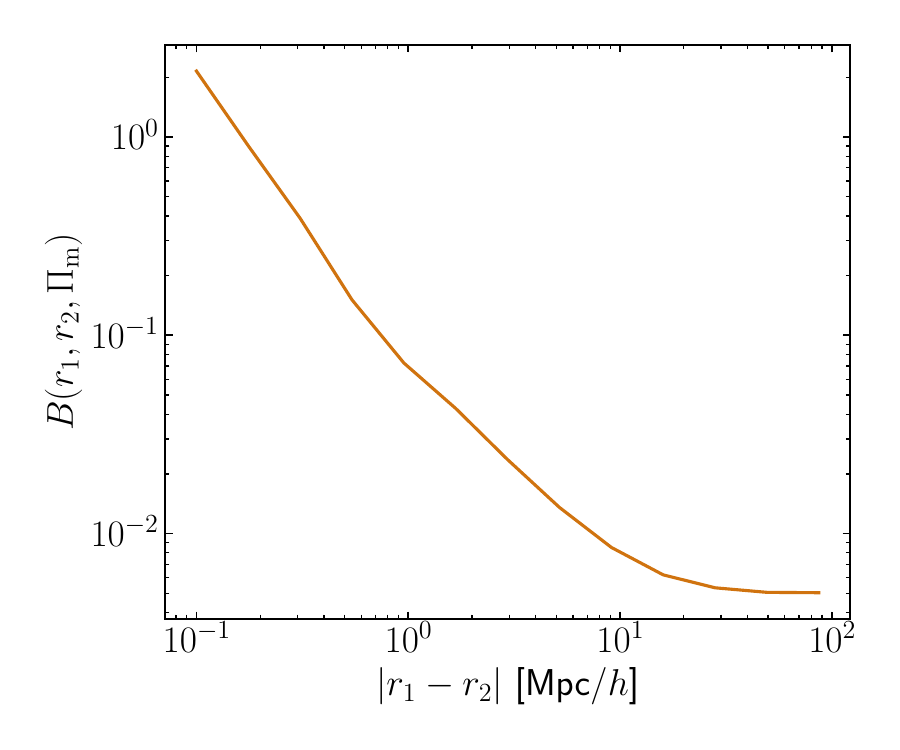}
    \caption{Estimator bias $B$ for the full LOWZ galaxy sample.}
    \label{fig:estbias}
\end{figure}

We show in Fig.~\ref{fig:estbias} the $B$ estimated for the full LOWZ sample. The bias is strongly scale-dependent. 

\end{appendix}

\end{document}